\documentclass[12pt]{revtex4-1}
\usepackage{amsmath,amssymb}
\usepackage{dcolumn}
\usepackage{bm}
\usepackage[usenames,dvipsnames]{xcolor}
\usepackage{amsthm,amscd}
\usepackage{graphicx}
\newcommand{\bib}[2]{\frac{\partial {#1}}{\partial {#2}}}

\def\w{{\wedge}}

\newcommand{\nc}{\newcommand}
\nc{\rnc}{\renewcommand}
\nc{\fr}[2]{\frac{#1}{#2}}
\nc{\eps}{\epsilon}
\rnc{\th}{\theta}
\nc{\vp}{\varphi}

\begin{document}
\title{
Energy-momentum conservation laws in Finsler/Kawaguchi Lagrangian formulation
}
\author{Takayoshi Ootsuka}
\email{ootsuka@cosmos.phys.ocha.ac.jp}
\affiliation{Physics Department, Ochanomizu University, 2-1-1 Ohtsuka Bunkyo-ku, Tokyo
112-8610, Japan}
\author{Muneyuki Ishida}
\email{ishida@phys.meisei-u.ac.jp}
\affiliation{Department of Physics, Meisei University,
 2-1-1 Hodokubo, Hino, Tokyo 191-8506, Japan }
\author{Erico Tanaka}
\email{erico@sci.kagoshima-u.ac.jp}
\affiliation{Department of Mathematics and Computer Science, 
Kagoshima University, 1-21-35 K\={o}rimoto Kagoshima, Kagoshima, Japan}
\author{Ryoko Yahagi}
\email{yahagi@hep.phys.ocha.ac.jp}
\affiliation{Physics Department, Ochanomizu University, 2-1-1 Ohtsuka Bunkyo-ku, Tokyo
112-8610, Japan}
\date{\today}

\begin{abstract}

We reformulate the standard Lagrangian formulation to a reparameterisation invariant
Lagrangian formulation by means of Finsler and Kawaguchi geometry.
In our formulation, various types of symmetries that appears in theories of physics 
are expressed geometrically by symmetries of Finsler (Kawaguchi) metric,
and the conservation law of energy-momentum arise as a part of Euler-Lagrange equations. 
The Euler-Lagrange equations are given geometrically in
reparameterisation invariant form,  
and the conserved energy-momentum 
currents can be obtained more easily, than by the conventional Lagrangian formulation.
The application to  scalar field, Dirac field, electromagnetic field
and general relativity are introduced. 
Especially, we propose an alternative definition 
of energy-momentum current of gravity, which satisfies gauge invariance under on-shell
condition.

\end{abstract}


\maketitle

\section{Introduction}

It is essential for an action integral to be defined independent of parameters 
so that the variational principle 
(the Hamilton's least action principle)
becomes a geometrical expression. 
Namely, the Lagrangian of the system needs to be reparameterisation invariant.
The standard way to derive the conservation law of the energy 
(energy-momentum current for field theory) 
is by the Noether's theorem in accord with the translational symmetry.
However, in the reparameterisation invariant system, 
it appears as a part of Euler-Lagrange equations. 

Let us take an example of a particle moving in the Schwarzschild spacetime:
\begin{eqnarray*}
 L(x^\mu,\dot{x}^\mu)=mc\sqrt{g_{\mu\nu}(x)\dot{x}^\mu \dot{x}^\nu},\quad
 g=\left(1-\frac{a}{r}\right)(dx^0)^2-\frac{(dr)^2}{1-a/r}
 -r^2\{(d\theta)^2+\sin^2{\theta}(d\varphi)^2\}.
\end{eqnarray*}
The Euler-Lagrange equations are given by,
\begin{eqnarray*}
\left\{
\begin{array}{l}
\displaystyle{ 0=\frac{d}{d\tau}\left(
\frac{mc g_{\mu 0}\dot{x}^\mu 
}{\sqrt{g_{\alpha\beta}\dot{x}^\alpha \dot{x}^\beta}}
\right), }\\
\displaystyle{ 0=
\left. {\frac{mc}2\bib{g_{\mu\nu}}{x^i}\dot{x}^\mu \dot{x}^\nu}
\middle/ {\sqrt{g_{\alpha\beta}\dot{x}^\alpha \dot{x}^\beta}} \right.
-\frac{d}{d\tau}\left(
\frac{mc g_{\mu i}\dot{x}^\mu 
}{\sqrt{g_{\alpha\beta}\dot{x}^\alpha \dot{x}^\beta}}
\right).}
\end{array}
\right.
\end{eqnarray*}
with , $i=1,2,3$. 
Notice that the first equation is what we call the energy conservation law of a 
relativistic particle.
This happens because the action of a relativistic particle is reparameterisation invariant, 
and only three equations out of four are independent. 
There is no reason that we should not choose this first equation as the equation of motion. 
The energy conservation law and the equations of motion are equivalent, in this sense.

We see the same mechanism in the model of a free bosonic string. 
The Lagrangian of Nambu-Goto action is,
\begin{eqnarray*}
L(X^\mu,{\dot{X}^\mu},{X}'^\mu)=
 \kappa_0 \sqrt{({\dot{X}_\mu} {X}'^\mu)^2
 -({\dot{X}}_\mu \dot{X}^\mu)  ({X}'_\nu {X}'^\nu)},
\end{eqnarray*}
with $\displaystyle{\dot{X}^\mu=\bib{X^\mu}{\tau},{X}'^\mu=\bib{X^\mu}{\sigma}}$.
The Euler-Lagrange equations are,
\begin{eqnarray*}
 0=\bib{}{\tau}\left\{
 \kappa_0
 \frac{
 ({X}')^2 \dot{X}_\mu - (\dot{X}\cdot {X}'){X}'_\mu
 }{\sqrt{(\dot{X}\cdot {X}')^2-(\dot{X})^2({X}')^2}}
 \right\}+
 \bib{}{\sigma}\left\{
 \kappa_0
 \frac{
 (\dot{X})^2 {X}'_\mu - (\dot{X}\cdot {X}')\dot{X}_\mu
 }{\sqrt{(\dot{X}\cdot {X}')^2-(\dot{X})^2({X}')^2}}
 \right\}, 
\end{eqnarray*}
with $\mu=0,1,\ldots, N$. 
These equations contain the conservation law of the energy-momentum current. 
We can see this by taking the spacetime parameters, $\tau=X^0, \, \sigma=X^1$, 
then the equation for $\mu=0$ ($\mu=1$) becomes the conservation law of energy (momentum) current.
This is also because the Nambu-Goto action is reparameterisation invariant.

However, these are specific results when the action is reparameterisation invariant, 
and without this invariance, 
such equations does not appear as Euler-Lagrange equations, even if it is a conserved system.
Nevertheless, it is known that any Lagrangian system of finite degrees of freedom 
can be rewritten in a reparameterisation invariant form 
without affecting its physical contents~\cite{Lanczos, Suzuki1956, OT2, Ootsuka2012, TaPHD2013}.

In this paper, we will further extend these results and show how to consider 
{\it every} Lagrangian systems of standard physical theory 
by the framework of reparameterisation invariant Lagrangian formulation. 
Conventionally, 
Lagrangian system is described by a set of configuration space and Lagrangian $(Q,L)$, 
but in general, this $(Q,L)$ is not a geometric space. 
In the reparameterisation invariant Lagrangian formulation, 
we will use the extended configuration space $M:=\mathbb{R}^{n+1} \times Q$
instead of $Q$, and Finsler metric $F$ (or Kawaguchi metric (areal metric) $K$ for field theory)
as a Lagrangian. 
The pair $(M,F)$ (for field theory, $(M,K)$) becomes a geometrical space; 
a space endowed with a {\it length (area)},
which is invariant under reparameterisation. 
The solution obtained by taking the variation of the action becomes an oriented curve 
(oriented $k$-dimensional submanifold) in the
Finsler (Kawaguchi) manifold. 
Since the action is given by taking the integral of the Finsler (Kawaguchi) metric
over the oriented curve (oriented $k$-dimensional submanifold), 
the Euler-Lagrange equations derived from this action are
apparently reparameterisation invariant, and therefore the energy (energy-momentum)
conservation law appears as their part. 
Thus, the previous examples could be reinterpreted as follows.

The first example, a relativistic particle moving in Schwarzschild spacetime 
is described by the Finsler manifold:
\begin{eqnarray*}
 M=\mathbb{R} \times \mathbb{R}_+ \times S^2, \quad F=m\sqrt{g_{\mu\nu}(x)dx^\mu dx^\nu},
\end{eqnarray*}
and the Nambu-Goto string is described by the Kawaguchi 
manifold~\cite{Ingarden1987}:
\begin{eqnarray*}
 M=\mathbb{R}^{N+1}, \quad K=\kappa_0 \sqrt{-\frac12 (dX_\mu \w dX_\nu)(dX^\mu \w dX^\nu)},
 \quad 
 \left( dX_{\mu}=\eta_{\mu\nu}dX^\nu  \right).
\end{eqnarray*}
As we mentioned, these are the special cases where the Lagrangian already 
had the property of reparameterisation invariance. 
However, our formulation is not restricted to such special cases, 
and we will later show some examples 
for the Lagrangian without this property.

In the next section, 
we will give the definition of Finsler and Kawaguchi manifold
used in our formulation. 
The Finsler-Kawaguchi Lagrangian formulation is described in section 3,
and the examples of a point particle, scalar field, Dirac field, and electromagnetic field
are introduced successively in section 4. 
We show that 
energy-momentum conservation law
in the standard sense 
appears in the Euler-Lagrange equations.  
In section 5, we apply the theory to general relativity and Palatini $f(R)$ gravity, 
and 
derive conserved currents naturally as in the same way introduced in section 4.
We propose that such quantities could be interpreted 
as energy-momentum currents of gravitational theories.

\section{Finsler and Kawaguchi manifold}

A Finsler manifold $(M,F)$ is a natural extension of a Riemannian manifold. 
$M$ is a differentiable manifold and the function $F$ defined by 
\begin{align}
 F:D(F)
 \subset TM \to \mathbb{R}, \qquad
 F:v \in D(F)\mapsto F(v) \in \mathbb{R}, \quad
 F(\lambda v)=\lambda F(v), \quad {}^\forall \lambda >0, \label{cond_hom}
\end{align}
is called the Finsler metric or the Finsler function
~\cite{Matsumoto,Bao-Chern-Shen,Bucataru-Miron}.
$D(F)$ is a sub-bundle of the tangent bundle $TM$ where the Finsler function is well-defined. 
Usually, in mathematical literatures, a slit tangent bundle $TM^{\circ}=TM\setminus\{0\}$ is
taken for this sub-bundle $D(F)$. 
However, from the viewpoint of physics, 
we need to consider it as a more general sub-bundle of $TM$, 
since it is not guaranteed that we could always have the function $F$ on the whole $TM^{\circ}$.
The last condition in (\ref{cond_hom}) is called the {\it homogeneity condition}.
The Finsler function gives a vector a geometrically well-defined norm, due to this condition.

In this paper, we formulate the application of Finsler geometry to a Lagrangian system
and derive its equations of motion and conserved currents.
Since the standard Lagrangians of physics are given in local coordinates, 
we will also give the definition of a Finsler manifold $(M,F)$ 
in local coordinates. 
Let $M$ be an $(n+1)$-dimensional differentiable manifold and $U$ be a subset of $M$.
The Finsler metric is written as a function of the coordinates $x^\mu$
and the 1-forms $dx^\mu$ with $(\mu=0,1,\ldots,n)$, on $U$. 
The latter $dx^\mu$ could be also regarded as 
adapted coordinates on $TM$. 
Throughout this paper, we have introduced definitions such as to make these 
two interpretations completely interchangable. 
In this way, several concepts such as prolongation which usually requires lengthy descriptions 
becomes quite simple. 
The homogeneity condition is expressed by, 
\begin{align}
 F\left(x^\mu,\lambda dx^\mu\right)=\lambda F\left(x^\mu,dx^\mu\right), \quad 
 {}^\forall \lambda >0.
 \label{homo}
\end{align}
The Finsler metric gives a tangent vector $\bm{v} \in D(F)_p \subset T_pM$ its norm by,
\begin{align}
 F\left(x^\mu(p),dx^\mu(\bm{v})\right)=F\left(x^\mu(p),v^\mu\right)\in \mathbb{R}.
\end{align}
Standard literatures of mathematics also assumes the following conditions: \\
i) (positivity) $F(v)>0$ \\
and \\
ii) (regularity) 
$\displaystyle g_{\mu\nu}(x,dx):=\frac12 \frac{\partial^2 F^2}{\partial dx^\mu \partial dx^\nu},
\,
{\rm det}\left(g_{\mu\nu}(x,dx)\right)\neq 0$. \\
However, for our motivation, these conditions are not necessary.
The only requirement for our theory is the homogeneity condition $(\ref{homo})$.

Next, we will define the Finsler length of an oriented curve $\bm{c}$ on $M$ by,
\begin{align}
 {\cal A}[\bm{c}]=\int_{\bm{c}}F:=\int_{s^0}^{s^1}
 F\left(x^\mu(s),\frac{dx^\mu(s)}{ds}\right)ds,
\end{align} 
where $c:[s^0,s^1]\to M$, is called a parameterisation and 
$x^\mu(s)=x^\mu(c(s))$C$\displaystyle{\frac{dx^\mu(s)}{ds}=\frac{dx^\mu(c(s))}{ds}}$. 
Notice that we used the same symbol $c$ for both submanifold and paramameterisation 
using bold face for the former. 
The pull back of $F=F\left(x^\mu,dx^\mu\right)$ by the map $c$ is naturally considered as
$c^\ast F:=F\left(c^\ast x^\mu,c^\ast dx^\mu\right)$, 
then the Finsler length ${\cal A}[\bm{c}]$ becomes an integration of a 1-form $c^\ast F$ 
over the interval $[s^0,s^1]$.
${\cal A}[\bm{c}]$ does not depend on its choice of parameter owing to the homogeneity condition 
$(\ref{homo})$.
In this sense, it is a well defined geometrical length for the oriented curve $\bm{c}$.

\bigskip

The field theory can be also formulated by the infinite dimensional Finsler manifold.
In this case, the theory is reparameterisation invariant only with respect to the ``time'' 
parameter.
However, we will show that the mathematical structure becomes more simple 
if we use the Kawaguchi manifold. 
It introduces us to a {\it finite dimensional} configuration space formulation.

A Kawaguchi manifold $(M,K)$ is a natural generalisation of Finsler manifold 
to a multi-dimensional parameter space. It is also called the 
$k$-dimensional areal space~\cite{Kawaguchi1964}. 
Here, $M$ is an $N$-dimensional differentiable manifold and $K$ is called the Kawaguchi metric.
$K$ defines a $k$-dimensional area for an oriented $k$-dimensional submanifold of $M$ ($1<k\leq N$).
We can construct its definition in parallel to Finsler geometry.
A Kawaguchi metric (or Kawaguchi function) $K$ is a function such that satisfies: 
\begin{align}
 K:D(K) \subset \Lambda^{k}TM \to \mathbb{R}, \qquad
 K:v^{[k]} \mapsto K(v^{[k]}), \qquad
 K(\lambda v^{[k]})=\lambda K(v^{[k]}), \quad {}^\forall \lambda >0, \label{cond_hom2}
\end{align}
where $D(K)$ is assumed to be a sub-bundle of $\Lambda^{k}TM$, 
and $\bm{v}^{[k]}=\frac{1}{k!}v^{\nu_1\nu_2\cdots \nu_k}\bib{}{x^{\nu_1}}\w 
\bib{}{x^{\nu_2}} \w \cdots \w \bib{}{x^{\nu_k}}
\in \Lambda^k T_pM$ is a $k$-vector, 
which express the $k$-dimensional oriented surface element at point $p\in M$. 
The last condition in (\ref{cond_hom2}) is called the homogeneity condition of Kawaguchi metric.
Again, since the usual field Lagrangians are given in coordinate expression, 
we also give the definition of a Kawaguchi manifold $(M,K)$ in local coordinates.

Let $x^\mu \, (\mu=1,\ldots, N)$ be the local coordinates of $M$.
We define the Kawaguchi metric as the function of $x^\mu$ and $k$-form
$dx^{\mu_{1}\mu_2\cdots \mu_k}:=dx^{\mu_1}\w dx^{\mu_2} \w
\cdots \w dx^{\mu_k}
\, (\mu_i=1,2,\ldots,N, \, i=1,2,\ldots,k)$.
The latter $k$-form could be also regarded as coordinate functions on $\Lambda^k TM$, 
but as in the case of Finsler, we opt to consider them as some variables on $M$, 
expressing the first-order derivatives. 
In these local coordinates, the homogeneity condition becomes,
\begin{align}
 K\left(x^\mu,\lambda dx^{\mu_1 \mu_2 \cdots \mu_k}\right)
 =\lambda K\left(x^\mu,dx^{\mu_1 \mu_2 \cdots \mu_k}\right), \quad 
 {}^\forall \lambda >0. \label{Khom}
\end{align}
As a generalisation of Finsler metric, Kawaguchi metric gives 
a geometric norm to a $k$-vector $\bm{v}^{[k]}$ by,
\begin{align}
 K\left(x^\mu(p),dx^{\mu_1\mu_2 \cdots \mu_k}(\bm{v}^{[k]})\right) 
 =K\left(x^\mu(p),v^{\mu_1\mu_2 \cdots \mu_k}\right) 
 \in \mathbb{R}, 
\end{align}
and by integration, defines the $k$-dimensional area to a 
$k$-dimensional oriented submanifold $\bm{\sigma}$ by: 
\begin{align}
 {\cal A}[\bm{\sigma}]=\int_{\bm{\sigma}}K
 :=\int_{W \subset \mathbb{R}^{k}} \hspace{-16pt}
 K\left(x^\mu(s^1,s^2,\cdots,s^k),
 \frac{\partial (x^{\mu_1},x^{\mu_2},\cdots,x^{\mu_k})}{\partial (s^1,s^2,\cdots,s^k)}\right)
 ds^1 \w ds^2 \w \cdots \w ds^k. \label{K-action}
\end{align}
Here, the map 
$\sigma:W \subset \mathbb{R}^k \to M$
is called a {\it parameterisation} of $\bm{\sigma}$, 
and the variables in (\ref{K-action}) are understood as,  
$x^\mu(s^1,s^2,\cdots,s^k)=x^\mu \left(\sigma(s^1,s^2,\cdots,s^k)\right)$C
and
\begin{align*}
\displaystyle{\frac{\partial (x^{\mu_1},x^{\mu_2},\cdots,x^{\mu_k})}
{\partial (s^1,s^2,\cdots,s^k)}}
:=\left|
\begin{array}{cccc}
 \bib{x^{\mu_1}}{s^1} & \bib{x^{\mu_1}}{s^2} & \cdots & \bib{x^{\mu_1}}{s^k} \\
 \bib{x^{\mu_2}}{s^1} & \bib{x^{\mu_2}}{s^2} & \cdots & \bib{x^{\mu_2}}{s^k} \\
 \vdots &&& \\
 \bib{x^{\mu_k}}{s^1} & \bib{x^{\mu_k}}{s^2} & \cdots & \bib{x^{\mu_k}}{s^k} 
\end{array}
\right|.
\end{align*}
We define the pull back of the Kawaguchi function $K$ by the map $\sigma$ as 
$\sigma^\ast K:=K\left(\sigma^\ast x^\mu,\sigma^\ast dx^{\mu_1 \mu_2 \cdots \mu_k}\right)$.
Then, by using the homogeneity condition, 
\begin{eqnarray*}
 \textstyle
 \sigma^\ast K
 &=&
 K\left(x^\mu(s^1,\cdots,s^k),\frac{\partial (x^{\mu_1},\cdots ,x^{\mu_k})}
 {\partial (s^1,\cdots,s^k)}ds^1 \w \cdots \w ds^k\right) \\
 &=&K\left(x^\mu(s^1,\cdots,s^k),\frac{\partial (x^{\mu_1},\cdots ,x^{\mu_k})}
 {\partial (s^1,\cdots,s^k)} \right) ds^1 \w \cdots \w ds^k
\end{eqnarray*}
becomes a $k$-form on $W$. 
Consequently, ${\cal A}[\bm{\sigma}]$ becomes a reparameterisation invariant area of 
$\bm{\sigma}$.

\section{Covariant Lagrangian formulation} 

Finsler geometry originated when considering the geometry of calculus of variations.
Therefore, it is a natural setting for formulating 
the variational principle considered in physics.

Firstly, we will explain how to handle the Lagrangian system with finite degrees of freedom
in terms of Finsler geometry.
It would be ideal if we could start from the definition of Finsler
manifold $(M,F)$ completely in a covariant fashion, namely, 
without any specific choice of $M$.
Physicists, however, always fix the ``time'' parameter during their experiments, 
and it is this physicist's view point we have to take into account. 
So, we will start our discussion with the pair of configuration space and Lagrangian: $(Q,L)$.
Note that this implies we have already selected a certain ``time'' parameter, 
and chose the theoretical model as $L\left(q^i,\dot{q}^i,t\right)$. 
We will construct our Finsler manifold $(M,F)$ in accord to this model $(Q,L)$, 
and it is given by the following~\cite{Lanczos, Suzuki1956}: 
\begin{align}
 M:=\mathbb{R}\times Q, \quad
 F\left(x^\mu,dx^\mu\right):=L\left(x^i,\frac{dx^i}{dx^0},x^0\right)dx^0,  \label{homotec1}
\end{align}
with $\mu=0,1,\cdots,n$, $i=1,2,\cdots,n$.
$M$ is the product space of time and configuration space $Q$,
and is called the extended configuration space.
It is easy to check that the above $F\left(x^\mu,dx^\mu\right)$ satisfies the 
homogeneity condition $(\ref{homo})$, and therefore is a Finsler metric. 
By the reparameterisation invariant property of Finsler metric, 
the choice of the ``time'' parameter does not affect its physical meaning. 
We will call this Finsler metric a {\it covariant Lagrangian} 
and our method a {\it covariant Lagrangian formulation}.

The trajectory of a point particle 
(an oriented curve $\bm{c}$ which satisfies the equations of motion) 
in the extended configuration space is determined by the principle of least action.
The action integral is given by ${\cal A}[\bm{c}]=\int_{\bm{c}}F$.

The Euler-Lagrange equations 
determine the extremal curve $\bm{c}$.
We set the initial point $p_0$ and the final point $p_1$ on $M$, and consider a differentiable map
$\varphi:[-\varepsilon_0,\varepsilon_1] \times M \to M$,
$\varphi(\varepsilon, \,):=\varphi_\varepsilon:M \to M$.
The map $\varphi_\varepsilon$ satisfies the conditions 
$\varphi_0=id_M$, $\varphi_\varepsilon(p_0)=p_0$, $\varphi_\varepsilon(p_1)=p_1$,
and $\varphi_{\varepsilon'}\circ \varphi_{\varepsilon}=\varphi_{\varepsilon'+\varepsilon}$, 
$(\varphi_{\varepsilon})^{-1}=\varphi_{-\varepsilon}$.
Such $\varphi$ is called a {\it flow} on $M$.
Let the vector field $X \in \Gamma (TM)$ be 
a generator: $\varphi_\varepsilon={\rm Exp}(\varepsilon X)$.
We define the variation of the curve by
$\displaystyle{\delta \bm{c}=\left.\frac{d}{d\varepsilon}\right|_{\varepsilon=0}
\varphi_\varepsilon(\bm{c})}$.
The principle of least action is described by
\begin{align}\label{eq_leastac}
 0=\delta {\cal A}[\bm{c}]
 &:=
 \left.\frac{d}{d\varepsilon}\right|_{\varepsilon=0}
 {\cal A}[\varphi_{\varepsilon}(\bm{c})]
 = \left.\frac{d}{d\varepsilon}\right|_{\varepsilon=0}
 \int_{\varphi_{\varepsilon}(\bm{c})} \hspace{-12pt} F
 = \left.\frac{d}{d\varepsilon}\right|_{\varepsilon=0}
 \int_{\bm{c}} \varphi_{\varepsilon}^\ast F.
\end{align}
Now, choose a parameterisation of the curve as $c:[s^0,s^1]\to M$, 
$c(s^0)=p_0, c(s^1)=p_1$.
Then (\ref{eq_leastac}) becomes,
\begin{align}\label{vars}
 \left.\frac{d}{d\varepsilon}\right|_{\varepsilon=0}
 \int_{\bm{c}}\varphi_\varepsilon^\ast F
 = \left.\frac{d}{d\varepsilon}\right|_{\varepsilon=0}
 \int_{s^0}^{s^1} c^\ast \varphi_\varepsilon^\ast F
 =
 \int_{s^0}^{s^1} 
 \left.\frac{d}{d\varepsilon}\right|_{\varepsilon=0}
 \hspace{-12pt}
 F\left(x^\mu(\varphi_{\varepsilon}(c(s))),
 dx^\mu(\varphi_{\varepsilon}(c(s)))\right). 
\end{align}
The integrand of the last part of $(\ref{vars})$ is evaluated as,
\begin{align}
 c^\ast \delta F&=\delta x^\mu c^\ast \left(\bib{F}{x^\mu}\right)+d\delta x^\mu 
 c^\ast \left(\bib{F}{dx^\mu}\right) \nonumber \\ 
 &=d\left[ \delta x^\mu c^\ast \left(\bib{F}{dx^\mu} \right) \right]
 +\delta x^\mu \left[ c^\ast \left(\bib{F}{x^\mu}\right)-d \left\{
 c^\ast \left(\bib{F}{dx^\mu}\right)\right\} \right]. \label{deltaF}
\end{align}
Here we used the notation $\displaystyle{\delta x^\mu
=\left.\frac{d}{d\varepsilon}\right|_{\varepsilon=0}x^\mu(\varphi_\varepsilon(c(s)))
=c^\ast {\cal L}_X x^\mu=c^\ast X^\mu}$, 
and ${\cal L}_X$ is the Lie derivative by the vector field $\displaystyle{X=X^\mu \bib{}{x^\mu}}$.
The term $\displaystyle{\bib{F}{dx^\mu}}$ is considered as a function of $x^\mu$ and $dx^\mu$,
so $\displaystyle{c^\ast \left(\bib{F}{dx^\mu}\right)
=\left(\bib{F}{dx^\mu}\right)(c^\ast x^\mu, c^\ast dx^\mu)}$.
In the calculation of $\int_{\bf c}\delta F$, the contribution from the
first term of $(\ref{deltaF})$ vanishes, because the vector field $X$ has
the condition $X(c(s^0))=X(c(s^1))=0$ at the end points.
For the other points, there are no restrictions for $\delta x^\mu=c^\ast X^\mu$. 
Therefore, the condition that the action is at its extremum becomes,  
\begin{align}
 0=c^\ast \left\{
 \bib{F}{x^\mu}-d\left(\bib{F}{dx^\mu}\right)
 \right\}, \quad (\mu=0,1,\ldots, n), \label{covEL1}
\end{align}
and such curve $\bm{c}$ is called the extremal of the action ${\cal A}[\bm{c}]$.
We call (\ref{covEL1}) the Euler-Lagrange equations.
These equations are reparameterisation invariant, since they are derived from 
a reparameterisation invariant
action integral.
Notice that the reparameterisation invariant property makes them dependent on each other.

We also comment on the Noether's theorem. 
Let us assume that the system has a certain symmetry. 
It is convenient to use the generalised expression of Lie derivative, 
and 
its action to the Finsler metric $F$ with respect to the vector field
$\displaystyle{v=v^\mu \bib{}{x^\mu}}$ is given by,
\begin{align}
 {\cal L}_v F:={\cal L}_v x^\mu \bib{F}{x^\mu}+{\cal L}_v dx^\mu \bib{F}{dx^\mu}
 =v^\mu \bib{F}{x^\mu}+dv^\mu \bib{F}{dx^\mu}. \label{LievF}
\end{align}
The vector field $v$ which satisfies ${\cal L}_v F=0$ 
is called the {\it symmetry} of $F$.
In other words, it is a Killing vector field of $F$.
This definition (\ref{LievF}) gives the same result as ${\cal L}_ {{\rm pr}^{(1)}v} F = {{\rm pr}^{(1)}v}(F)$, 
when $dx^\mu$ is interpreted as the adapted coordinates of $TM$, 
and the prolongation of $v$ to $TM$ is given by,   
\begin{align}
 {\rm pr}^{(1)}v=v^\mu \bib{}{x^\mu}+dv^\mu\bib{}{dx^\mu}.
\end{align}
We call this a covariant prolongation of $v$, or a vertical lift of $v$. 
It is possible to compare this to the standard expression 
using jet bundle formulation~\cite{Olver}. 
Considering $x^0$ as a parameter, we have
\begin{align}
 {\cal L}_v F&=v^0\bib{F}{x^0}+v^a\bib{F}{x^a}
 +dv^0\bib{F}{dx^0}+dv^a\bib{F}{dx^a} \nonumber  \\
 &=v^0 \bib{F}{x^0}+v^a\bib{F}{x^a}+\frac{dv^0}{dx^0}
 \left(F-\bib{F}{dx^a}dx^a\right)+dv^a\bib{F}{dx^a} \nonumber  \\
 &=v^0 \bib{F}{x^0}+v^a\bib{F}{x^a}
 +\left(\frac{dv^a}{dx^0}-\frac{dv^0}{dx^0}\frac{dx^a}{dx^0}\right)\bib{F}{\dot{x}^a}
 +\frac{dv^0}{dx^0}F,  \quad (a=1,2,\dots,n)
\end{align}
where we used the homogeneity condition
\begin{align}
 \bib{F}{dx^\mu}dx^\mu=\bib{F}{dx^0}dx^0+\bib{F}{dx^a}dx^a=F,
\end{align}
to obtain second equality, then with $\displaystyle{\dot{x}^a:=\frac{dx^a}{dx^0}}$ and the relation
$\displaystyle{\bib{}{dx^a}=\bib{\dot{x}^b}{dx^a}\bib{}{\dot{x}^b}=\frac{1}{dx^0}\bib{}{\dot{x}^a}}$, 
obtained the third equality. 
Rewriting the coordinates by the notation following~\cite{Olver}, {\em i.e.},
\begin{align}
 t:=x^0, \quad q^a:=x^a, \quad \xi:=v^0, \quad \phi^a:=v^a, \quad
 \dot{q}^a:=\frac{dx^a}{dx^0}, \quad
 \bib{}{t}:=\bib{}{x^0}, \quad \bib{}{\dot{q}^a}=\bib{}{\dot{x}^a}, \nonumber
\end{align}
and taking $\displaystyle{F=L\left(x^0,x^a,\frac{dx^a}{dx^0}\right)dx^0}$,
we obtain the equation
\begin{align}
\tilde{pr}^{(1)}v(L)+L Div(\xi)=0, \label{varsym}
\end{align}
where 
\begin{align} 
 \tilde{pr}^{(1)}v= 
 \xi \bib{}{t}+\phi^a \bib{}{q^a}
 +\left(\frac{d\phi^a}{dt}-\frac{d\xi}{dt}\dot{q}^a\right)
 \bib{}{\dot{q}^a} , \quad
Div(\xi)= \frac{d\xi}{dt}.
\end{align}
The group of transformation satisfying equation (\ref{varsym}) is called a {\em variational symmetry} in \cite{Olver}, 
and $\tilde{pr}^{(1)}v$ is the prolongation with respect to the bundle structure. 
The equation (\ref{varsym}) is equivalent to our ${\cal L}_v F=0$, but  
notice that our expression is much simpler, due to the fact that we consider no fibration.
In practice, considering ${\cal L}_v$ as the standard Lie derivative and calculation such as
\begin{eqnarray}
 {\cal L}_v dx^\mu=d\left({\cal L}_v x^\mu\right)=dv^\mu, \quad
 {\cal L}_v F=\left({\cal L}_v x^\mu\right) \bib{F}{x^\mu}
 +\left({\cal L}_v dx^\mu\right)\bib{F}{dx^\mu},
\end{eqnarray}
is sufficient to obtain the same result as in the conventional framework. 
From the point of calculational efficiency, this is also a notable result. 

Considering 
\begin{align*}
\displaystyle{{\cal L}_v F = d\left[v^\mu \frac{\partial F}{\partial dx^\mu}\right]
+v^\mu \left\{ \frac{\partial F}{\partial x^\mu}
-d\left( \frac{\partial F}{\partial dx^\mu}\right)\right\}},
\end{align*}
under the assumption that Euler-Lagrange equations are satisfied; namely, 
the curve $\bm{c}$ is the extremal, we obtain a conservation law: 
\begin{align}
 c^\ast d\left[v^\mu \left(\bib{F}{dx^\mu}\right)\right] = 0, \label{eq_noether}
\end{align}
This is the expression of the Noether's theorem by our formalism.

If the Lagrangian $L$ does not contain $x^0$ explicitly ({\it i.e.} a conserved system),
the Finsler metric constructed by $(\ref{homotec1})$ also does not include $x^0$.
In this case, $x^0$ is called a {\it cyclic coordinate}, and its Euler-Lagrange equation
for $\mu=0$ represents the energy conservation law of this system.
On the other hand, the conservation law is also obtained directly 
by inserting the generator $\displaystyle{v=\bib{}{x^0}}$ to (\ref{eq_noether}).
Either way leads to the same expression.

\bigskip

Secondly, we will move on to the field theory, 
that is the Lagrangian system with infinite degrees of freedom.
As we have mentioned in the beginning of this section, 
it would be better if we could start from the definition 
of the Kawaguchi manifold $(M,F)$, with general $M$.  
However, under normal circumstances, 
we can only observe the nature by fixing the ``spacetime'', 
namely the parameter space $W$, 
as we have fixed the ``time'' parameter for the case of dynamical systems.
Therefore, we will start by considering the standard Lagrangian system 
$(E\stackrel{\pi}{\to}W,Q,L)$, 
where $E \stackrel{\pi}{\to} W, \, Q$ are the vector bundle and its fibre
~\cite{Binz-Sniatycki-Fischer, Olver}.
We choose the total space $E$ to be our Kawaguchi manifold $M$, 
$\mathrm{dim} M=\mathrm{dim} W+\mathrm{dim} Q$.
The Kawaguchi metric $K$ is constructed from the Lagrangian 
$\displaystyle{L\left(u^A,\bib{u^A}{x^\mu}\right)}$
as follows~\cite{Ootsuka2012, TaPHD2013},
\begin{align}
 K\left(z^a,dz^{abcd}\right)=
 L\left(u^A,
 \frac{\varepsilon_{\mu\nu\rho\sigma}}{3!}
 \frac{dx^{\nu\rho\sigma} \w du^A}{dx^{0123}}
\right)dx^{0123}. \label{Kdef}
\end{align}
Here, 
$(z^a):=(x^\mu,u^A), a=0,1,\ldots,D+3, \mu=0,1,2,3, A=1,2,\ldots,D$,
where $D$ is the degree of freedom of fields.
The totally anti-symmetric Levi-Civita symbol
$\varepsilon_{\mu\nu\rho\sigma}, \, (\mu,\nu,\rho,\sigma=0,1,2,3)$
has the convention: $\varepsilon_{0123}=-1$.
Note that the field variables: $u^A$ are treated as independent variables, 
just as the spacetime coordinates $x^\mu$ are.
This is the major difference from the standard Lagrangian formulation.
The $K$ constructed in this way satisfies the homogeneity condition $(\ref{Khom})$, 
and we obtain our Kawaguchi manifold, $(M,K)$. 
The second argument of $(\ref{Kdef})$ may look a little complicated, 
nevertheless, its pull back with respect to the spacetime parameters, namely $x^\mu$, 
gives the standard variables, $\displaystyle{\bib{u^A}{x^\mu}}$. 
The action integral is given by ${\cal A}[\bm{\sigma}]=\int_{\bm{\sigma}}K$,
where $\bm{\sigma}$ is a $4$-dimensional oriented submanifold in $M$.
As before, the least action principle is described by the map
$\varphi_\varepsilon={\rm Exp}(\varepsilon X)$ on $M$ which is fixed on the boundary.
Then the variation of $K$ by $X$ becomes, 
\begin{align}
 \sigma^\ast \delta K&=\delta z^a \sigma^\ast \left(\bib{K}{z^a}\right)
 +\frac{1}{3!}d\delta z^a \w dz^{bcd} \sigma^\ast \left(\bib{K}{dz^{abcd}}\right) \nonumber \\
 &=d\left[
 \delta z^a \sigma^\ast \left(\frac1{3!}\bib{K}{dz^{abcd}}dz^{bcd}\right)
 \right]
 +\delta z^a \left[
 \sigma^\ast \left(\bib{K}{z^a}\right)-d\left\{
 \frac{1}{3!}\sigma^\ast \left(\bib{K}{dz^{abcd}}dz^{bcd}\right)
 \right\}
 \right], \label{delta-K}
\end{align}
where we had taken arbitrary spacetime parameterisation $\sigma:W \to \bm{\sigma} \subset M$.
Next, we set $z^a_\varepsilon(s):=z^a(\varphi_\varepsilon(\sigma(s)))$,
and differentiate $\sigma^\ast \varphi_\varepsilon^\ast K
=K(z^a_\varepsilon(s), dz^a_\varepsilon(s) \w dz^b_\varepsilon(s) \linebreak[3]
\w dz^c_\varepsilon(s) \linebreak[3]
\w dz^d_\varepsilon(s))$ with respect to $\varepsilon$. 
By similar considerations as in the case of Finsler, 
we obtain the Euler-Lagrange field equations,
\begin{align}
 0=\sigma^\ast \left\{
 \bib{K}{z^a}-
 d\left( \frac{1}{3!}\bib{K}{dz^{abcd}} dz^{bcd}
 \right)\right\}. \label{covEL2}
\end{align}
These equations are reparameterisation invariant, 
and again, at least four of them are dependent on each other. 

The Noether's theorem could be also obtained for the field theory.   
The expression of generalised Lie derivative of the Kawaguchi metric $K$ 
with respect to the vector field 
$\displaystyle{v=v^a\bib{}{z^a}}$
on $M$ is now given by 
\begin{align}
 {\cal L}_v K:={\cal L}_v dz^a \bib{K}{z^a}+{\cal L}_v dz^{abcd} \frac1{4!}\bib{K}{dz^{abcd}}
 =v^a \bib{K}{z^a}+\frac1{3!}dv^a \w dz^{bcd} \bib{K}{dz^{abcd}}.
\end{align}
The vector field $v$ such that satisfies ${\cal L}_v K=0$ 
is called the {\it symmetry} of $K$,
or the Killing vector field of $K$.
Under the condition that the system satisfies the Euler-Lagrange equations,
we obtain
\begin{align}
 \sigma^\ast d\left[v^a \left(
 \frac1{3!} \bib{K}{dz^{abcd}} dz^{bcd}\right)\right]=0, \label{eq_conserv}
\end{align}
as a conservation law.

If the Lagrangian $L$ does not include the coordinates $(x^0,x^1,x^2,x^3)$ explicitly,
they become cyclic coordinates, and the equations $(\ref{covEL2})$ for
$a=0, 1, 2, 3,$ will represent the conservation law of energy-momentum.
The same conservation law could be also derived by inserting the Killing vector 
$\displaystyle{v=\bib{}{z^a}, a=0, 1, 2, 3}$
to (\ref{eq_conserv}).

\section{Examples}

From this section, we will omit the pull back symbol $c^\ast$ ($\sigma^\ast$ for Kawaguchi)
unless we need to emphasize, for notational simpilicity. 
However, it is important to keep in mind that these equations hold only 
on the submanifold, $\bm{c}$ ($\bm{\sigma}$).

\subsection{Newtonian mechanics} 

We begin with an example of Newtonian mechanics, 
using the Lagrangian formulation of Finsler geometry.
Let $L$ be the Lagrangian of a potential system for an $n$-dimensional space:
$\displaystyle{L=\sum_{i=1}^n\frac{m}{2}(\dot{q}^i)^2-V(q^1,q^2,\cdots,q^n)}$.
Here, $m$ is the mass of the particle.
We define the Finsler manifold $(M,F)$ by,
\begin{align}
 M=\{(x^0,x^1,\cdots,x^n)\}\simeq \mathbb{R}^{n+1}, \quad
 F(x^\mu,dx^\mu)=\sum_{i=1}^n \frac{m (dx^i)^2}{2dx^0}-V(x^1,\cdots,x^n)dx^0. 
\end{align}
Note that this $F$ is defined 
only on the sub-bundle $D(F)=TM \setminus \{dx^0=0\}$.
The Euler-Lagrange equations become, 
\begin{align}
 0&= -d \left(\frac{\partial F}{\partial dx^0}\right)
 = d\left[
 \sum_{i=1}^n \frac{m}{2} \left(\frac{dx^i}{dx^0}\right)^2+V(x^i)\right], 
 \label{L0}\\
 0&= \frac{\partial F}{\partial x^i}-d \left(
 \frac{\partial F}{\partial dx^i} \right)
 = -\bib{V}{x^i}dx^0-d\left(m\frac{dx^i}{dx^0}\right), \quad (i=1,2,\ldots,n).
 \label{L1}
\end{align}
The reparameterisation invariance gives us the freedom to choose the time parameter
$s$, $c:[s^0,s^1]\to M$.
The standard choice is to take $s=x^0$, that is, 
$c^\ast x^0 = s$, $c^\ast dx^0=ds$,
$c^\ast x^i=x^i(s)$, $\displaystyle{c^\ast dx^i=\frac{dx^i(s)}{ds} ds}$,
and one can verify that (\ref{L0}), (\ref{L1}) gives the conventional conservation law of energy 
and equations of motion.
However, from the perspective of the covariant Finsler formulation, 
such choice of parameterisation is not obligatory, 
and we may take a parameterisation such as $s=x^1$, 
under the assumption we are only considering on the local coordinate system. 
This is one of the significant results of our formalism.

The conservation law $(\ref{L0})$ can be also derived from the Noether's theorem, namely, 
\begin{align}
 {\cal L}_{\bib{}{x^0}} F=\bib{F}{x^0}=0 \quad \Rightarrow \quad d\left(\bib{F}{dx^0}\right)=0.
\end{align}

\subsection{Scalar field theory} 

The first example of field theory is the real scalar field theory
on $4$-dimensional Minkowski spacetime $(\mathbb{R}^4,\eta)$.
In local coordinate system, 
$\eta=\eta_{\mu\nu}dx^\mu \otimes dx^\nu$,
$\eta_{00}=-\eta_{11}=-\eta_{22}=-\eta_{33}=1$ and $\eta_{\mu\nu}=0, \, (\mu\neq \nu)$.
The conventional Lagrangian is $L=\frac12 \partial^\mu \phi \partial_\mu \phi-V(\phi)$, 
where $V(\phi)$ is the potential term.
The Kawaguchi manifold obtained from this Lagrangian becomes
\begin{align}
 M=\{(x^\mu,\phi)\} \simeq \mathbb{R}^4 \times \mathbb{R}, \quad
 K =-\frac{(dx_{\mu\nu\rho}\w d\phi)(dx^{\mu\nu\rho} \w d\phi)}{2\cdot 3!dx^{0123}}
  -V(\phi)dx^{0123},\label{sK}
\end{align}
$M$ is the extended configuration space, 
and we use abbreviations and notations such as 
$dx^{\mu\nu\rho}:=dx^\mu \w dx^\nu \w dx^\rho$,
$dx_{\mu}:=\eta_{\mu\nu}dx^\nu$.
By $(\ref{sK})$,
$D(K)=\Lambda^4 TM \setminus\{dx^{0123}=0\}$.
The Euler-Lagrange equations are derived by using $(\ref{covEL2})$, 
\begin{align}
 0&=d\left[
 \frac{dx_{\mu\nu\rho} \w d\phi}{2!dx^{0123}}dx^{\nu\rho}\w d\phi
 -\left\{
 -\frac{(dx_{\alpha\beta\gamma} \w d\phi)(dx^{\alpha\beta\gamma} \w d\phi)}
 {2\cdot 3!(dx^{0123})^2}
 +V(\phi)
 \right\}\frac{1}{3!}\varepsilon_{\mu\nu\rho\sigma}dx^{\nu\rho\sigma}
 \right], \label{sEL0} \\
 0&=-V'(\phi)dx^{0123}+d\left\{-
 \frac{dx_{\mu\nu\rho}\w d\phi}{3! dx^{0123}} dx^{\mu\nu\rho}
 \right\}. \label{sEL1}
\end{align}
It is also possible to derive these equations 
by directly calculating the variation, $(\ref{delta-K})$. 
Usually, for more complex systems, 
the calculation is more simple by the latter method. 
We can naturally define a energy-momentum current in the covariant form,
\begin{align}
 \tilde{J}_\mu:= \frac{dx_{\mu\nu\rho} \w d\phi}{2!dx^{0123}}dx^{\nu\rho}\w d\phi
 -\left\{
 -\frac{(dx_{\alpha\beta\gamma} \w d\phi)(dx^{\alpha\beta\gamma} \w d\phi)}
 {2\cdot 3!(dx^{0123})^2}
 +V(\phi)
 \right\}\frac{1}{3!}\varepsilon_{\mu\nu\rho\sigma}dx^{\nu\rho\sigma},  
 \label{sEM}
\end{align}
for $\mu=0,1,2,3$.
To avoid confusion, 
the quantities on the Kawaguchi manifolds are denoted with a tilde $\tilde{}$.
The four equation of motion $(\ref{sEL0})$ indicates 
that these currents are conserved, namely $d\tilde{J}_\mu=0$.
This means $d \left(\sigma^\ast \tilde{J}_\mu \right)=0$
for arbitrary spacetime parameterisation $\sigma$.

As in the previous example, 
the coordinates $x^\mu, \, (\mu=0,1,2,3)$ are cyclic coordinates, 
and therefore it is possible to see the conservation law 
directly as a part of Euler-Lagrange equations.

Now we will look into the details of this simple example of scalar field theory. 
From our point of view, the conventional theory in the framework of Minkowski spacetime 
corresponds to the case where a specific parameterisation is chosen in the set up 
of Kawaguchi manifold. 
We rewrite the coordinate functions of Kawaguchi spacetime 
as $z^a, (a=0,1,\ldots, 4)$, 
where $(z^a):=(x^\mu,\phi)$.
The ordinary
choice of parameterisation $\sigma$ is expressed by 
$\sigma(x):W \subset \mathbb{R}^4 \to M, \, 
\sigma^\ast z^\mu=x^\mu, \, \sigma^\ast z^4=\phi(x)$. 
This means that we are simply taking the coordinates of Minkowski spacetime as parameters.
The pull back of the Kawaguchi metric to the parameter space becomes,
\begin{eqnarray*}
 \sigma(x)^\ast K
 &=& 
 -\frac{(dx_{\mu\nu\rho} \w dx^\alpha~\partial_\alpha \phi) 
  (dx^{\mu\nu\rho\beta}~\partial_\beta \phi)}{2 \cdot 3! dx^{0123}}-V(\phi)dx^{0123} \\
 &=&\left\{ -\frac{
 {\varepsilon_{\mu\nu\rho}}^\alpha~\varepsilon^{\mu\nu\rho\beta}~
 \partial_\alpha \phi~\partial_\beta \phi}{2\cdot 3!}
 -V(\phi)
 \right\}dx^{0123}=\left\{
 \frac12 \partial^\mu \phi~\partial_\mu \phi-V(\phi)\right\}dx^{0123},
\end{eqnarray*}
which is just the conventional Lagrangian function times the volume form 
of Minkowski spacetime.
The second equality is obtained by the cancelation of $dx^{0123}$ 
which appears by the pull back on the numerator.

Next, we will also pull back the Euler-Lagrange equations 
by this specific parameterisation, $\sigma(x)$. 
Consider $\phi(x)$ as a function of $x^\mu$, and treating $d$ as an exterior derivative, 
we get, $\displaystyle{\frac{dx_{123} \w d\phi}{dx^{0123}}dx^{123}
=-\frac{dx^{1230}\partial_0 \phi}{dx^{0123}}dx^{123}=\partial_0 \phi~dx^{123},}$
therefore, the pull back of $(\ref{sEL1})$ by $\sigma(x)$ becomes,
\begin{eqnarray*}
 0&=&-V'(\phi)dx^{0123}+d\left(-\partial_0 \phi~dx^{123}-\partial_1 \phi~dx^{023}
 -\partial_2 \phi~dx^{031}-\partial_3 \phi~dx^{012}\right) \\
 &=& \left\{-V'(\phi)-\partial_0^2 \phi+\partial_1^2 \phi
 +\partial_2^2 \phi+\partial_3^2 \phi\right\}dx^{0123},
\end{eqnarray*}
which is the standard wave equation of $\phi$. 
Similarly, the pull back of energy-momentum current $(\ref{sEM})$ for $\mu=0,1$ becomes,
\begin{eqnarray*}
 J_0&=&
 \left(
 \partial_1 \phi~dx^{23}+\partial_2 \phi~dx^{31}+\partial_3 \phi~dx^{12}
 \right)\w d\phi
 +\left\{
 {\textstyle \frac12} \partial^\mu \phi \partial_\mu \phi+V(\phi)
 \right\}dx^{123} \\
 &=&
 \left\{\textstyle
 \frac{(\partial_0 \phi)^2+(\partial_1 \phi)^2+(\partial_2 \phi)^2
 +(\partial_3 \phi)^2}{2}
 +V(\phi)\right\}dx^{123}
 +\partial_0 \phi \partial_1 \phi dx^{023}
 +\partial_0 \phi \partial_2 \phi dx^{031}
 +\partial_0 \phi \partial_3 \phi dx^{012},
 \\
 J_1&=&
 \left(
 \partial_0 \phi~dx^{23}-\partial_3 \phi~dx^{02}+\partial_2 \phi~dx^{03}
 \right)\w d\phi
 -\left\{
 {\textstyle \frac12} \partial^\mu \phi \partial_\mu \phi+V(\phi)
 \right\}dx^{023} \\
 &=&
 \partial_0 \phi \partial_1 \phi dx^{123}
 +\left\{\textstyle
 \frac{(\partial_0 \phi)^2+(\partial_1 \phi)^2-(\partial_2 \phi)^2
 -(\partial_3 \phi)^2}{2}
 -V(\phi)\right\}dx^{023}
 +\partial_1 \phi \partial_2 \phi dx^{031}
 +\partial_1 \phi \partial_3 \phi dx^{012},
\end{eqnarray*}
which is also the well-known definition of the standard energy-momentum current. 

A well-established approach to deal field theory by means of geometry is 
to use a fibre bundle (normally a vector bundle) structure, 
where the base manifold is the $4$-dimensional spacetime, 
and the fields are described by the section of the bundle. 
The theory formulated on such structure does not depend on the coordinates of the 
base manifold, meaning that we can use arbitrary spacetime coordinates 
$f^\mu({x}^\nu)$ as spacetime parameters. 
This is the standard meaning of covariance. 
On the contrary, we have formulated the field theory on a Kawaguchi manifold, 
without any reference to fibred structures.  
In such approach, the spacetime coordinates and field variables are treated equally as 
coordinate functions of the Kawaguchi manifold, and coordinate transformations of the type   
$\tilde{f}^\mu(x^\nu,\phi)$, $\phi$ denoting the field, 
does not change the theory. 
We call such a property, {\it an extended covariance}. 

To see this more clearly, first consider a free relativistic particle moving in 
a Minkowski spacetime $(M,\eta)$. 
By special relativity, we know that there is no specific 
time coordinate for $M$, and this means we can always choose an appropriate time parameter to 
describe the trajectory of the particle. 
The geodesic is obtained as an extremal of the action, by performing the calculus of variation. 
In this case, $(M,F),F=\sqrt{\eta_{\mu\nu}dx^\mu dx^\nu}$ is the Finsler manifold where 
we constructed the Lagrangian formulation, $F$ is the Lagrangian 
and the geodesic is a submanifold of $M$. 
On the other hand, the non-relativistic description of the free particle is to consider a fibre bundle 
where the base space $U$ is the space of time parameter, usually a subset of $\mathbb{R}$, 
and total space is a direct product, i.e., $M=U \times \mathbb{R}^3$. 
In this case, the geodesic of the particle is given by a section of this fibre bundle.  
In such sense, the general covariance is a property arising from the deletion of the bundle structure. 
The construction of Lagrangian formulation on Kawaguchi manifold $(M,K)$ 
does the same for the case of field theory. 
We have removed the fibre bundle structure of the standard field theory 
(where the base space was given by $U\in \mathbb{R}^4$ and total space by $M=U\times \Sigma$), 
and instead of considering the field configuration as its section, 
gave it as a $4$-dimensional submanifold of the total space $M$. 
Performing calculus of variation {\it determines} the spacetime, as the extremal. 
The extended covariance is a property obtained by deleting the bundle structure. 

\subsection{Dirac field theory} 

The next example is the theory of free Dirac field. The conventional Lagrangian is given by 
$L=\frac{i}2\left(\bar{\psi}\gamma^\mu \partial_\mu \psi-\partial_\mu \bar{\psi}
\gamma^\mu \psi \right)-m\bar{\psi}\psi$,
where $\psi$ is a spinor, and $\bar{\psi}:=\psi^\dagger \gamma^0$ is its Dirac conjugate. 
We also supressed the indices, such as $\psi=(\psi^A)$,
$\bar{\psi}=\psi^\dagger \gamma^0=(\bar{\psi}_A)$,
$\gamma^\mu \psi=({(\gamma^\mu)^B}_A \psi^A)$. 
The Kawaguchi manifold becomes,
\begin{align}
 M&=\{(x^\mu,\psi,\bar{\psi})\} \simeq \mathbb{R}^4 \times \mathbb{C}^4,
 \nonumber \\
 K&=\frac{1}{2\cdot 3!}\left(
 \bar{\psi} \gamma^5 \gamma_{\mu\nu\rho} dx^{\mu\nu\rho} 
 \w d\psi
 - d\bar{\psi}\w dx^{\mu\nu\rho} 
 \gamma_{\mu\nu\rho} \gamma^5  \psi 
 \right) 
 -m\bar{\psi}\psi dx^{0123}, \label{Dirac_K}
\end{align}
with the convention $\gamma^5=i\gamma^0 \gamma^1 \gamma^2 \gamma^3$ and
$\gamma_{\mu\nu\rho}=\gamma_{[\mu}\gamma_{\nu}\gamma_{\rho]}$ 
($\gamma_{012}=\gamma_0 \gamma_1 \gamma_2$, $\gamma_{011}=0$ etc.).
The Euler-Lagrange equations are derived by using $(\ref{covEL2})$,
\begin{align}
 0&=d\left(
 -\frac{
 \bar{\psi} \gamma^5 \gamma_{\mu\nu\sigma} dx^{\mu\nu} \w d\psi
 +d\bar{\psi} \w dx^{\mu\nu} \gamma_{\mu\nu\sigma} \gamma^5 \psi
 }{2 \cdot 2!}
 +\frac{1}{3!}\varepsilon_{\mu\nu\rho\sigma} m\bar{\psi}\psi dx^{\mu\nu\rho} 
 \right), \label{DEL0}\\
 0&=\frac{1}{3!} \gamma^5 \gamma_{\mu\nu\rho} dx^{\mu\nu\rho} \w d\psi-m\psi dx^{0123}, \\
 0&= -\frac{1}{3!} d\bar{\psi} \w dx^{\mu\nu\rho} \gamma_{\mu\nu\rho} \gamma^5
 -m\bar{\psi}dx^{0123}.
\end{align}
Since spinors are Grassmann variables, 
note that differentiation with respect to $\psi$ ($\bar{\psi}$) must be taken 
by the right (left) derivatives. 
The equation $(\ref{DEL0})$ 
are the conservation laws of energy-momentum currents.
As in the previous examples, the coordinates $x^\mu, \, (\mu=0,1,2,3)$ are cyclic coordinates, 
and this is the reason we can see the conservation law directly 
as a part of Euler-Lagrange equations.
Similar discussions will follow for the choice of arbitrary parameters and 
the relation to the conventional theory.

\subsection{Electromagnetic field theory} 

From the conventional Lagrangian of free electromagnetic field:  $L=-\frac14 F^{\mu\nu}F_{\mu\nu}$,
we obtain our Kawaguchi manifold as,
\begin{align}
 M
 =\{(x^\mu,A_\mu)\}\simeq \mathbb{R}^8, \quad
 K
 =\frac{(\tilde{F} \w dx_{\rho\sigma})(\tilde{F} \w dx^{\rho\sigma})}
 {4dx^{0123}}, \quad (dx^{0123}\neq 0),
\end{align}
where $\tilde{F}=dA_\mu \w dx^\mu$.
The Euler-Lagrange equations are derived as,
\begin{align}
 0&=d\left\{
 \frac{\tilde{F} \w dx_{\rho\sigma}}{dx^{0123}} \tilde{F} \w dx^\rho
 +\varepsilon_{\mu\nu\rho\sigma}
 \frac{(\tilde{F} \w dx_{\alpha\beta})(\tilde{F} \w dx^{\alpha\beta})}{4 \cdot 3!(dx^{0123})^2}
 dx^{\mu\nu\rho}
 +\frac{\tilde{F} \w dx_{\mu\nu}}{2dx^{0123}} dA_\sigma \w dx^{\mu\nu}
 \right\}, \label{Maxwell0} \\
 0&=d\left(\frac{\tilde{F} \w dx_{\rho\sigma}}{2dx^{0123}}dx^{\mu\rho\sigma}\right).
 \label{Maxwell1}
\end{align}
Equation $(\ref{Maxwell0})$ represents the conservation law of energy-momentum current 
of electromagnetic field, 
and we can define the current by 
\begin{align} 
 \tilde{J}_\mu= \frac{\tilde{F} \w dx_{\rho\mu}}{dx^{0123}} \tilde{F} \w dx^\rho
 -\varepsilon_{\mu\nu\rho\sigma}
 \frac{(\tilde{F} \w dx_{\alpha\beta})(\tilde{F} \w dx^{\alpha\beta})}{4 \cdot 3!(dx^{0123})^2}
 dx^{\nu\rho\sigma}
 +\frac{\tilde{F} \w dx_{\rho\sigma}}{2dx^{0123}} dA_\mu \w dx^{\rho\sigma}.
\end{align}
The pull back of the equations $(\ref{Maxwell0})$ and $(\ref{Maxwell1})$ to the parameter space
by $\sigma(x)$ is:  
\begin{align}
 0&=d\left(
 -\frac14 {\varepsilon^{\rho\sigma}}_{\mu\nu}
 F_{\rho\sigma} F_{\alpha\beta} dx^{\alpha\beta\nu}
 +\frac1{4\cdot 3!} \varepsilon_{\mu\nu\rho\sigma} F_{\alpha\beta}F^{\alpha\beta}
 dx^{\nu\rho\sigma}
 +\frac14 {\varepsilon^{\alpha\beta}}_{\rho\sigma} F_{\alpha\beta}
 dA_\mu \w dx^{\rho\sigma}\right) , \label{eq_maxwell_current1} \\
 0&=-\partial_\nu F^{\mu\nu}dx^{0123}. \label{eq_maxwell_current2} 
\end{align}
The last term of the pull backed current (\ref{eq_maxwell_current1}) 
is not gauge invariant with respect to the 
usual gauge transformation $A_\rho\rightarrow A_\rho+\partial_\rho\chi$. 
However, by using (\ref{eq_maxwell_current2}), this term becomes an exact term,
\begin{eqnarray*}
 \frac14{\varepsilon^{\alpha\beta}}_{\rho\sigma} F_{\alpha\beta}
 dA_\mu \w dx^{\rho\sigma}
 =d\left( \frac14
 {\varepsilon^{\alpha\beta}}_{\rho\sigma}A_\mu F_{\alpha\beta}
 dx^{\rho\sigma} 
 \right).
\end{eqnarray*}

\subsection{Maxwell-Dirac field theory}

Here we will combine the last two examples, and 
consider the Dirac field interacting with the electromagnetic field.
The Kawaguchi manifold is given by,
\begin{align}
 M=\{(x^\mu,A_\mu,\psi,\bar{\psi})\}\simeq \mathbb{R}^8 \times \mathbb{C}^4, \quad
 K=K_{\rm Maxwell}+K_{\rm Dirac},
\end{align}
where
\begin{align}
 &K_{\rm Maxwell}:=\frac{(\tilde{F} \w dx_{\rho\sigma})
 (\tilde{F}\w dx^{\rho\sigma})}{4dx^{0123}}, \quad (dx^{0123}\neq 0),\\
 &K_{\rm Dirac}:=
 \frac{1}{2 \cdot 3!}\left(
 \bar{\psi} \gamma^5 \gamma_{\mu\nu\rho} dx^{\mu\nu\rho} \w D\psi -
 \bar{D}\bar{\psi} \w dx^{\mu\nu\rho} \gamma_{\mu\nu\rho} \gamma^5 \psi 
 \right)
 -m\bar{\psi}\psi dx^{0123}.
\end{align}
The covariant derivatives are defined by
$D\psi=d\psi-ieA_\mu dx^\mu \psi$ and $\bar{D}\bar{\psi}=d\bar{\psi}+ieA_\mu dx^\mu \bar{\psi}$.

The Euler-Lagrange equations become, 
\begin{align}
 &0=
 d\left\{
 -\frac{\tilde{F}\w dx_{\mu\rho}}{dx^{0123}} \tilde{F} \w dx^\rho
 -\varepsilon_{\mu\nu\rho\sigma}
 \frac{(\tilde{F} \w dx_{\alpha\beta})(\tilde{F} \w dx^{\alpha\beta})}{4\cdot3!(dx^{0123})^2} 
 dx^{\nu\rho\sigma}
 +\frac{\tilde{F} \w dx_{\rho\sigma}}{2dx^{0123}} dA_\mu \w dx^{\rho\sigma}
 \right. \label{Maxwell-Dirac0}  \\
 &
 \left.
 -\frac{\bar{\psi} \gamma^5 \gamma_{\mu\nu\rho} dx^{\nu\rho} \w D\psi
 +\bar{D}\bar{\psi} \w dx^{\nu\rho} \gamma_{\mu\nu\rho} \gamma^5 \psi}{2\cdot 2!}
 -\frac{1}{3!}\varepsilon_{\mu\nu\rho\sigma}m\bar{\psi}\psi dx^{\nu\rho\sigma}
 +\frac{1}{3!}ie \bar{\psi} \gamma_{\nu\rho\sigma} \gamma^5 A_\mu \psi dx^{\nu\rho\sigma} 
 \right\},
 \label{Maxwell-Dirac1}\\
 &0=
 \frac{1}{3!}ie \bar{\psi} \gamma^5 \gamma_{\nu\rho\sigma} dx^{\mu\nu\rho\sigma} \psi
 -d\left\{\frac{\tilde{F} \w dx_{\rho\sigma}}{2 dx^{0123}} dx^{\mu\rho\sigma}
 \right\},
 \label{Maxwell-Dirac2} \\
 &0=\frac1{3!}\gamma^5 \gamma_{\mu\nu\rho} dx^{\mu\nu\rho} \w D\psi -m\psi dx^{0123}, \\
 &0=-\frac1{3!} \bar{D}\bar{\psi} \w dx^{\mu\nu\rho} \gamma_{\mu\nu\rho} \gamma^5
 -m\bar{\psi}dx^{0123}. 
\end{align}
The equation (\ref{Maxwell-Dirac0}) expresses the energy-momentum conservation law 
of Maxwell-Dirac field theory. 
This Kawaguchi metric has a gauge symmetry described by the vector field, 
\begin{eqnarray}
 {\cal G}=\frac{\overleftarrow{\partial}}{\partial \psi}(ie\Lambda \psi) 
 -ie\bar{\psi} \Lambda \frac{\overrightarrow{\partial}}{\partial \bar{\psi}}
 +\bib{\Lambda}{x^\mu}\frac{{\partial}}{\partial A_\mu},
\end{eqnarray}
where, $\Lambda=\Lambda(x^\mu)$ is an arbitrary function of $x^\mu$.
The corresponding transformation is the usual gauge transformation we are familiar with:
$\delta\psi (= {\cal L}_{\cal G} \psi) = ie\Lambda \psi $,
$\delta\bar{\psi} = -ie\bar{\psi} \Lambda$,
$\displaystyle{\delta A_\mu =\bib{\Lambda}{x^\mu}}$,
$\delta x^\mu =0$,
$\delta D\psi = ie\Lambda D\psi$,
$\delta \bar{D}\bar{\psi} = -ie\Lambda \bar{D}\bar{\psi}$,
and $\delta \tilde{F} =0$.
One can check the condition ${\cal L}_{\cal G}K=0$ easily.
Taking the variation of the Kawaguchi metric by the vector field ${\cal G}$ 
under on-shell conditions
generates a conserved current:
\begin{eqnarray}
 \tilde{J}_{\cal G}&:=&\frac{{\cal L}_{\cal G}\bar{\psi}\gamma^5 \gamma_{\mu\nu\rho}\psi
 +\bar{\psi}\gamma_{\mu\nu\rho}\gamma^5 {\cal L}_{\cal G}\psi}
 {2 \cdot 3!}dx^{\mu\nu\rho}
 +{\cal L}_{\cal G}A_\mu \frac{\tilde{F} \w dx_{\rho\sigma}}{2 dx^{0123}}dx^{\mu\rho\sigma} 
\nonumber \\
 &=&ie \Lambda \left(
 \bar{\psi} \gamma_{\mu\nu\rho} \gamma^5 \psi
 \right) \frac{1}{3!}dx^{\mu\nu\rho}
 +\bib{\Lambda}{x^\mu}
 \frac{\tilde{F} \w dx_{\rho\sigma}}{2 dx^{0123}}dx^{\mu\rho\sigma}.
\end{eqnarray}
Its exterior derivative becomes,
\begin{align}
0=d\tilde{J}_{\cal G}=\Lambda d\left\{
 ie \frac{\bar{\psi} \gamma_{\mu\nu\rho} \gamma^5 \psi}{3!}dx^{\mu\nu\rho}
 \right\}+\bib{\Lambda}{x^\mu}
 \left\{
 ie \frac{\bar{\psi} \gamma_{\nu\rho\sigma} \gamma^5 \psi}{3!}dx^{\mu\nu\rho\sigma}
 +d \left(
 \frac{\tilde{F} \w dx_{\rho\sigma}}{2 dx^{0123}}dx^{\mu\rho\sigma}
 \right)
 \right\}.
\end{align}
This is the Noether's theorem. 
Since the functions $\Lambda$ and $\displaystyle{\bib{\Lambda}{x^\sigma}}$ are arbitrary,
we have the electric charge conservation law, 
\begin{eqnarray}
 d\tilde{J}_{e}=0, \quad \tilde{J}_e=-ie  \frac{\bar{\psi} \gamma^5 \gamma_{\mu\nu\rho} \psi}
 {3!}dx^{\mu\nu\rho},
\end{eqnarray}
and Maxwell equations $(\ref{Maxwell-Dirac2})$.

\section{Application to general relativity} 

Application to the Hilbert action of Einstein's general relativity requires
a more generalised Kawaguchi manifold; {\it higher-derivative areal space},
since the action includes second order derivatives.
It takes two steps to define higher-derivative areal space:
higher-derivative extension of Finsler metric, and areal extension of the former.

Higher-derivative Kawaguchi metric defines the length of the oriented curve 
on manifold $M$ as a function of higher-order derivatives.
Second order Finsler metric, which is usually called {\it Kawaguchi metric},
$F(x^\mu,dx^\mu,d^2x^\mu)$, for instance,
is defined as a function of $x^\mu, \, dx^\mu, \, d^2x^\mu \, (\mu=0,1,2,\dots,n)$
which satisfies the 
second order
homogeneity condition,
\begin{align}
 F(x^\mu,\lambda dx^\mu,\lambda^2 d^2 x^\mu+\xi dx^\mu)=\lambda F(x^\mu,dx^\mu,d^2 x^\mu),
 \quad {}^\forall \lambda >0, \, {}^\forall \xi \in \mathbb{R}. \label{highhom}
\end{align}
Differentiation of the above condition with respect to $\lambda$ and $\xi$ gives
\begin{align}
 &
 \bib{F}{dx^\mu}dx^\mu
 + 2\bib{F}{d^2 x^\mu}d^2 x^\mu
 =F, \\
 &\bib{F}{d^2 x^\mu}dx^\mu=0,
\end{align}
after setting $\lambda=1$, $\xi=0$, respectively.
This is the the well-known Zermelo's condition~\cite{Kawaguchi1976, Miron, Urban}.

The condition (\ref{highhom}) implies the Zermelo's condition. 
A parameterisation $c(s):[s_0,s_1] \to \bm{c} \subset M$ of the oriented curve $\bm{c}$
determines the pull back of $x^\mu,dx^\mu,d^2 x^\mu$ as 
\begin{align}
 &
 \displaystyle
 c(s)^\ast x^\mu=x^\mu(c(s)):=x^\mu(s), \quad
 c(s)^\ast dx^\mu=dx^\mu(s)=\frac{dx^\mu(s)}{ds} ds, \\ 
 &
 \displaystyle
 c(s)^\ast d^2 x^\mu
 :=d\left(\frac{dx^\mu(s)}{ds}ds\right)
 =d\left(\frac{dx^\mu(s)}{ds}\right)ds
 +\left(\frac{dx^\mu(s)}{ds}\right)d^2s
 := \frac{d^2 x^\mu(s)}{ds^2}ds^2,
\end{align}
and the pull back of $F$ is given by
\begin{align}
c^\ast F:=F\left(c^\ast x^\mu, c^\ast dx^\mu, c^\ast d^2 x^\mu \right)
 =F \left(x^\mu(s),\frac{dx^\mu(s)}{ds},
 \frac{d^2 x^\mu(s)}{ds^2}\right)ds.
\end{align}
The length of the oriented curve $\bm{c}$ is then defined by
\begin{align}
 {\cal A}[\bm{c}]=\int_{s_0}^{s_1} F\left(x^\mu(s),\frac{dx^\mu(s)}{ds},
 \frac{d^2 x^\mu(s)}{ds^2}\right)ds. \label{highlen}
\end{align}
In fact, another parameterisation $\tilde{c}(t):[t_0,t_1]\to \bm{c}$ yields
\begin{align}
\int_{t_0}^{t_1} F\left(
 x^\mu(t), \frac{dx^\mu(t)}{dt},\frac{d^2x^\mu(t)}{dt^2}
 \right) dt=\int_{t_0}^{t_1}
 F\left(
 x^\mu(t), \frac{ds}{dt}\frac{dx^\mu(s)}{ds},
 \left(\frac{ds}{dt}\right)^2\frac{d^2x^\mu(s)}{ds^2}+\frac{d^2s}{dt^2}
 \frac{dx^\mu(s)}{ds}
 \right) dt,
\end{align}
which corresponds to (\ref{highlen}), thanks to the homogeneity condition (\ref{highhom}).

Second order Kawaguchi metric (second order areal metric) 
$K\left(z^a,dz^{abcd},dz^{efg} \w d^2z^{abcd} \right)$
is a function of $z^a$, $dz^{abcd}$ and $dz^{efg} \w d^2 z^{abcd}:=dz^{efg} \w d(dz^{abcd})$, 
where $z^a$ are coordinate functions of a differentiable manifold $M$.
The last argument expresses the second order derivatives by our notation. 
In order to keep the reparameterisation invariance, it is necessary that 
the Kawaguchi metric satisfy the following condition,
\begin{align}
 K\left(z^a,\lambda dz^{abcd},
 \lambda^2 dz^{efg} \w d^2z^{abcd}
 +\mu^{efg} dz^{abcd}\right)
 =\lambda K\left(z^a,dz^{abcd},dz^{efg}\w d^2z^{abcd}\right),
{}^\forall \lambda >0, {}^\forall \mu^{efg} \in \mathbb{R}. \label{homo3}
\end{align}
Here, $\lambda >0$ is the arbitrary constant also appeared in the the case of 
first order homogeneity condition, 
and $\mu^{efg}$ are the constants in accord to the second order derivatives. 
They are completely antisymmetric in the superscripts.

We call the pair $(M,K)$ a 
{\it second order Kawaguchi manifold}, and (\ref{homo3}) the 
{\it second order homogeneity condition}. 
The second order homogeneity condition guarantees 
the Kawaguchi manifold the important property of 
reparameterisation invariance. 

From (\ref{homo3}), we can obtain the generalised version of 
Zermelo's condition to areal spaces simply 
by differentiating the equtaion (\ref{homo3}) 
with respect to $\lambda$ and $\mu^{efg}$ and then setting 
$\lambda=1$, $\mu^{efg}=0$; 
\begin{align}
 &
 \frac{1}{4!}
 \bib{K}{dz^{abcd}}dz^{abcd}
 +\frac{2}{3!4!} \bib{K}{dz^{efg}\w d^2 z^{abcd}}
 dz^{efg}\w d^2 z^{abcd}
 =K, \\
 &\bib{K}{dz^{efg}\w d^2 z^{abcd}}dz^{abcd}=0.
\end{align}

Let $\bm{\sigma}$ be an oriented $4$-dimensional submanifold embedded in $M$,
and its parameterisation given by 
$\sigma_0(s^0,s^1,s^2,s^3):W_0\subset \mathbb{R}^4 \to \bm{\sigma} \subset M$. 
Our second order variable $dz^{efg}\w d^2 z^{abcd}$ 
is related to the standard second order derivative by the pull back of $\sigma_0$ 
defined by, 
\begin{align}
 \sigma_0^\ast \left(
 dz^{efg}\w d^2 z^{abcd}
 \right)
 :=\frac{\partial\left(z^e, z^f, z^g,
  \frac{\partial(z^a,z^b,z^c,z^d)}{\partial(s^0,s^1,s^2,s^3)}
  \right)}{\partial(s^0,s^1,s^2,s^3)} 
 \left(ds^{0123}\right)^2.
\end{align}
Now, let $\sigma_1(t^0,t^1,t^2,t^3):W_1\subset \mathbb{R}^4 \to \bm{\sigma}$ 
be another parameterisation of $\bm{\sigma}$, 
and suppose we have an orientation preserving diffeomorphism $f:W_1 \to W_0$, such that 
$\sigma_1 = \sigma_0 \circ f$.
Then, the pull back of 
$\sigma_0^\ast \left(dz^{efg}\w d^2 z^{abcd}\right)$ by $f$ becomes, 
\begin{align}
 &f^\ast \circ \sigma_0^\ast \left(  dz^{efg}\w d^2 z^{abcd} \right)
 =\frac{\partial\left(z^e, z^f, z^g,
  \frac{\partial(z^a,z^b,z^c,z^d)}{\partial(t^0,t^1,t^2,t^3)}
  \frac{\partial(t^0,t^1,t^2,t^3)}{\partial(s^0,s^1,s^2,s^3)}
  \right)}{\partial(t^0,t^1,t^2,t^3)}
  \frac{\partial(s^0,s^1,s^2,s^3)}{\partial(t^0,t^1,t^2,t^3)} 
 \left(dt^{0123}\right)^2 \nonumber \\
 &\qquad =
  \left(dt^{0123}\right)^2 
 \frac{\partial\left(z^e, z^f, z^g,
  \frac{\partial(z^a,z^b,z^c,z^d)}{\partial(t^0,t^1,t^2,t^3)}
  \right)}{\partial(t^0,t^1,t^2,t^3)}
 \nonumber \\
 &\qquad + \left(dt^{0123}\right)^2  \frac{\partial\left(z^e, z^f, z^g,
  \frac{\partial(t^0,t^1,t^2,t^3)}{\partial(s^0,s^1,s^2,s^3)}
  \right)}{\partial(t^0,t^1,t^2,t^3)}
  \frac{\partial(s^0,s^1,s^2,s^3)}{\partial(t^0,t^1,t^2,t^3)} 
   \frac{\partial(z^a,z^b,z^c,z^d)}{\partial(t^0,t^1,t^2,t^3)}.
 \label{fsigma0}
\end{align}
The {\it r.h.s.} is equal to 
$\sigma_1^\ast  \left(dz^{efg}\w d^2 z^{abcd} +\mu^{efg}dz^{abcd}\right)$, {\it i.e.}, 
the standard relation $\sigma_1^\ast = f^\ast \circ \sigma_0^\ast$ does not hold for this variable,
due to the non-linearity of $dz^{efg}\w d^2 z^{abcd}$.
Next, we will define the pull back of second order Kawaguchi metric by 
$\sigma_0(s)$ such that, 
$\sigma_0^\ast K:=K\left(\sigma_0^\ast z^a,\sigma_0^\ast dz^{abcd},
 \sigma_0^\ast dz^{efg} \w d^2z^{abcd} \right).$
This is a $4$-form on $W_0$. We will further pull back this $\sigma_0^\ast K$ 
to a $4$-form on $W_1$ by $f$.
By considering the homogeneity condition $(\ref{homo3})$ of $K$ and the relation $(\ref{fsigma0})$, 
we find,
\begin{align}
 f^\ast \circ \sigma_0(s)^\ast K=\sigma_1(t)^\ast K,
\end{align}
despite the non-linearity of the second order variables. 
This property indicates that, as in the case of Finsler or first order Kawaguchi metric, 
the integration of the 
second order Kawaguchi metric
$K$ over $\bm{\sigma}$ 
also gives a reparameterisation invariant area for an oriented 
$4$-dimensional submanifold of $M$: 
\begin{align}
 {\cal A}[\bm{\sigma}]=\int_{\bm{\sigma}} K
 :=\int_{W} K\left(\sigma^\ast z^a,\sigma^\ast\left(dz^{abcd}\right),
 \sigma^\ast \left(dz^{efg} \w d^2z^{abcd} \right)\right). 
\end{align}

Suppose we are given the usual Lagrangian of second order field theory,
$\displaystyle{L\left(u^A, \bib{u^A}{x^\mu}, \linebreak[3] 
\frac{\partial^2 u^A}{\partial x^\mu \partial x^\nu}\right)}$,
then we can construct the second order Kawaguchi metric by,  
\begin{align}
 &K\left(z^a,dz^{abcd},dz^{efg} \w d^2z^{abcd}\right) \nonumber \\
 &=L\left(u^A,
 \frac{\varepsilon_{\mu\alpha\beta\gamma}}{3!}
 \frac{dx^{\alpha\beta\gamma}\w du^A}{dx^{0123}},
  \frac{\varepsilon_{\nu\xi\eta\zeta}}{3!}{dx^{\xi\eta\zeta}\w d
 \left(
  \frac{\varepsilon_{\mu\alpha\beta\gamma}}{3!}
 \frac{dx^{\alpha\beta\gamma}\w du^A}{dx^{0123}}
 \right)}\middle/{dx^{0123}}
 \right)dx^{0123}, \label{rel_2ndK-L}
\end{align}
where $\varepsilon^{0123}=1,\varepsilon_{0123}=-1$,and $(z^a)=(x^\mu,u^A)$.
The second order variable in (\ref{rel_2ndK-L}) is a short expression: 
\begin{align}
 dx^{\xi\eta\zeta} \w d\left(\frac{dx^{\alpha\beta\gamma}\w du^A}{dx^{0123}}\right)
 :=\frac{dx^{\xi\eta\zeta} \w d\left(dx^{\alpha\beta\gamma}\w du^A\right)(dx^{0123})
 -\left(dx^{\alpha\beta\gamma}\w du^A \right) dx^{\xi\eta\zeta} \w 
 d^2 x^{0123}
 }{(dx^{0123})^2}. \nonumber
\end{align}
One can check that the Kawaguchi metric constructed 
in this way satisfies the homogeneity condition $(\ref{homo3})$, 
and together with $M=\{(x^\mu,u^A)\}$, we obtain the second order Kawaguchi manifold, $(M,K)$. 

The Lagrangian of the 
vacuum general relativity 
with cosmological constant $\lambda$ is given by,
\begin{align}
 L=\sqrt{-g}\left(
 -\frac{r}{2\kappa}-\frac{\lambda}{\kappa}
 \right),
\end{align}
where $\displaystyle{\kappa=\frac{8\pi G}{c^4}}$, $R_{\mu\nu}={R^{\alpha}}_{\mu\alpha\nu}$,
$r=g^{\mu\nu}R_{\mu\nu}={R^{\mu\nu}}_{\mu\nu}$, with all Greek indices running from $0$ to $3$.
The Kawaguchi manifold $(M,K)$ constructed from this Lagrangian is, 
\begin{align}
 &M=\{(x^\mu, g^{\mu\nu})\}=\{(z^a)\}\simeq \mathbb{R}^{14}, \\
 &K\left(z^a,dz^{abcd},dz^{efg} \w d^2z^{abcd} \right)
 =\frac1{4\kappa} \varepsilon_{\mu\nu\rho\sigma}\sqrt{-g}~
 \tilde{R}^{\mu\nu} \w dx^{\rho\sigma}
 -\frac{\lambda}{\kappa} \sqrt{-g}~dx^{0123}, \label{KEmetric} \\
& \tilde{R}^{\mu\nu}:=g^{\nu\xi}\tilde{R}^\mu{}_\xi,
 \quad \tilde{R}^{\mu}{}_\xi:
 =d\tilde{\varGamma}^{\mu}{}_\xi+\tilde{\varGamma}^\mu{}_\lambda
 \w \tilde{\varGamma}^{\lambda}{}_\xi, \quad
 \tilde{\varGamma}^\mu {}_\xi:=g^{\mu\zeta}\tilde{\varGamma}_{\zeta\xi\eta}dx^\eta,
 \label{R} \\
& \tilde{\varGamma}_{\zeta\xi\eta} :=\frac12 
 \left(
  \varepsilon_{\xi\alpha\beta\gamma}
  \frac{dx^{\alpha\beta\gamma} \w dg_{\zeta\eta}}{3! dx^{0123}}
 +\varepsilon_{\eta\alpha\beta\gamma}
  \frac{dx^{\alpha\beta\gamma}\w dg_{\xi\zeta}}{3! dx^{0123}}
 -\varepsilon_{\zeta\alpha\beta\gamma}
  \frac{dx^{\alpha\beta\gamma} \w dg_{\xi\eta}}{3! dx^{0123}}
 \right). \label{connection}
\end{align}
Latin indices runs from $0$ to $13$, and if we use the unified coordinate system $\{(z^a)\}$, 
$(dz^{abcd})$ denotes $\left(dx^{0123}, dx^{\alpha\beta\gamma}\w dg^{\mu\nu}\right)$, 
and 
$\left(dz^{efg} \w d^2z^{abcd}\right)$ denotes 
$\left(dx^{\rho\sigma\zeta} \w d^2 x^{0123},
dx^{\rho\sigma\zeta} \w d(dx^{\alpha\beta\gamma}\w dg^{\mu\nu})\right)$. 

Here we emphasize that in our framework of Kawaguchi manifold, 
the variable $g^{\mu\nu}$ which conventionally correspond to the metric of Riemannian manifold, 
is merely treated as a coordinate function, similarly as the spacetime coordinates $x^\mu$. 
Each $10$ components of the symmetric matrix $g^{\mu \nu}$ represents independent coordinate function, 
and the variable $g_{\mu\nu}$ is the inverse of this symmetric matrix $g^{\mu\nu}$. 
In this sense, $g^{\mu \nu}$ does not represent any geometric structure. 
The Kawaguchi metric is the only geometrical structure we need. 

Before proceeding, let us check if this Kawaguchi metric is a plausible one. 
We pull back $K$ by the spacetime parameterisation 
$\sigma(x)$, 
which we used to verify the case of scalar field theory. 
The pull back by 
$\sigma(x)$
actually corresponds to considering the variables $g^{\mu\nu}$ as 
dependent variables of $x^\mu$. 
In this way, the pull back of $(\ref{R})$ 
becomes the usual curvature tensor, 
$\sigma(x)^\ast \tilde{R}^{\mu\nu}=\frac12 {R^{\mu\nu}}_{\alpha\beta}dx^{\alpha\beta}$,
and the Kawaguchi metric becomes,
\begin{align}
 \sigma^\ast K=\sqrt{-g}\left(
 -\frac{r}{2\kappa}-\frac{\lambda}{\kappa}
 \right)dx^{0123},
\end{align}
which is the standard Einstein-Hilbert Lagrangian $4$-form.

The general expressions of Euler-Lagrange equations can be obtained by 
considering the variational principle. 
However, in some cases, it is much more easier to directly take the variation of the 
concrete Kawaguchi action, and we will take this approach.
Remember, that in the covariant Lagrangian formulation, 
taking the variation $\delta$ means to take the 
Lie derivative with respect to arbitrary $X\in\Gamma(TM)$, 
and Lie derivative is commutative with $d$. 
For visibility, we will omit the pull back symbol $\sigma^\ast$ in the following discussion.  

The variation of $K$ becomes,
\begin{align}
 \delta K&=\frac1{4\kappa} \sqrt{-g}\left(
 -\frac12 \varepsilon_{\mu\nu\rho\sigma} g_{\xi\eta} \tilde{R}^{\mu\nu} \w dx^{\rho\sigma} 
 +\varepsilon_{\mu\eta\rho\sigma}\tilde{R}^\mu{}_\xi \w dx^{\rho\sigma}
 +2 \lambda g_{\xi\eta} 
 dx^{0123}\right) \delta g^{\xi\eta} \nonumber \\
 &
 +d\left(
 \frac1{4\kappa} \varepsilon_{\mu\nu\rho\sigma}\sqrt{-g}g^{\nu\xi}
 \delta \tilde{\varGamma}^\mu{}_\xi 
 \w dx^{\rho\sigma}
 \right) \nonumber \\
 &
 +\delta \tilde{\varGamma}^\mu {}_{\xi} \w 
 \left[
 \frac1{4\kappa} 
 \left\{
 \varepsilon_{\mu\nu\rho\sigma} 
  d\left(\sqrt{-g}g^{\nu\xi} dx^{\rho\sigma}\right)
 + \sqrt{-g}
 \left(
 \varepsilon_{\mu\nu\rho\sigma}
 g^{\nu\eta}\tilde{\varGamma}^\xi{}_\eta 
 -\varepsilon_{\eta\nu\rho\sigma}
 g^{\nu\xi}\tilde{\varGamma}^\eta{}_\mu 
 \right) \w dx^{\rho\sigma}
 \right\}
\right]
 \nonumber \\
 &
 -d\left\{
 \frac{1}{2\kappa}
 \varepsilon_{\mu\nu\rho\sigma}\sqrt{-g}\left(
 \tilde{R}^{\mu\nu} \w dx^\rho
 +\frac{2\lambda}{3!} dx^{\mu\nu\rho} 
 \right) \delta x^\sigma 
 \right\} \nonumber \\
 &
 +d\left\{
 \frac{1}{2\kappa}
 \varepsilon_{\mu\nu\rho\sigma}\sqrt{-g}\left(
 \tilde{R}^{\mu\nu} \w dx^\rho
 +\frac{2\lambda}{3!} dx^{\mu\nu\rho} 
 \right) \right\} \delta x^\sigma.
\label{dK}
\end{align}

The Euler-Lagrange equations described by the pull back of 
the parameterisation $\sigma:W\subset \mathbb{R}^4 \to M$ are the conditions for $4$-dimensional
submanifold $\bm{\sigma}$ to be an extremal submanifold of ${\cal A}[\bm{\sigma}]$.
We can use the following conditions to simplify the terms of $\delta K$:
\begin{align}
 \tilde{\varGamma}^{\mu}{}_{\rho\nu}= \tilde{\varGamma}^{\mu}{}_{\nu\rho}, \quad 
 dg_{\mu\nu}-g_{\xi\nu}\tilde{\varGamma}^\xi{}_\mu-g_{\mu\xi}\tilde{\varGamma}^\xi{}_\nu
 \stackrel{\sigma}{=}0,\label{dgGamma}
\end{align}
where the sign $\stackrel{\sigma}{=}$ 
means the equality on the $4$-dimensional submanifold 
${\bm \sigma}$ embedded in $M$ (ref. Appendix A),
and the second 
equality holds by,
\begin{align}
 &g_{\xi\nu}\tilde{\varGamma}^\xi{}_\mu+g_{\mu\xi}\tilde{\varGamma}^\xi{}_\nu
 =\tilde{\varGamma}_{\mu\nu}+\tilde{\varGamma}_{\nu\mu}
 =\left(
  \tilde{\varGamma}_{\mu\nu\rho}+\tilde{\varGamma}_{\nu\mu\rho} 
 \right)dx^\rho 
 =\frac{\varepsilon_{\rho\alpha\beta\gamma}}{3!}
  \frac{dx^{\alpha\beta\gamma} \w dg_{\mu\nu}}{dx^{0123}}dx^\rho \nonumber \\
 &\stackrel{\sigma}{=}-\frac{\varepsilon_{\rho\alpha\beta\gamma}}{3!}
 \left(
   \frac{dx^{\beta\gamma} \w dg_{\mu\nu} \w dx^\rho}{dx^{0123}}   dx^\alpha
 + \frac{dx^{\gamma} \w dg_{\mu\nu} \w dx^{\rho\alpha}}{dx^{0123}} dx^\beta
 + \frac{dg_{\mu\nu} \w dx^{\rho\alpha\beta}}{dx^{0123}}          dx^\gamma
 + \frac{dx^{\rho\alpha\beta\gamma}}{dx^{0123}}                   dg_{\mu\nu}
 \right) \nonumber \\
 &= -3\frac{\varepsilon_{\rho\alpha\beta\gamma}}{3!}
  \frac{dx^{\alpha\beta\gamma} \w dg_{\mu\nu}}{dx^{0123}}dx^\rho
 -\frac{\varepsilon_{\rho\alpha\beta\gamma}}{3!}\frac{dx^{\rho\alpha\beta\gamma}}{dx^{0123}}
 dg_{\mu\nu} \nonumber \\
 &\stackrel{\sigma}{=} - \frac{\varepsilon_{\rho\alpha\beta\gamma}}{4!}
 \frac{dx^{\rho\alpha\beta\gamma}}{dx^{0123}} dg_{\mu\nu}
 \stackrel{\sigma}{=} dg_{\mu\nu}. 
\end{align}
The term $\delta \tilde{\varGamma}^\mu{}_{\xi}$ in Eq.~(\ref{dK}) 
becomes zero under these conditions since,
\begin{align}
 &
 \varepsilon_{\mu\nu\rho\sigma} d\left(\sqrt{-g}g^{\nu\xi} dx^{\rho\sigma}\right)
 + \sqrt{-g}\left(
 \varepsilon_{\mu\nu\rho\sigma}
 g^{\nu\eta}\tilde{\varGamma}^\xi{}_\eta
 -\varepsilon_{\eta\nu\rho\sigma}
 g^{\nu\xi}\tilde{\varGamma}^\eta{}_\mu 
 \right) \w dx^{\rho\sigma} \nonumber \\
 &=\varepsilon_{\mu\nu\rho\sigma}\sqrt{-g}
 \left(\frac{1}{2}g^{\nu\xi}g^{\alpha\beta}dg_{\alpha\beta}
 -g^{\alpha\nu}dg_{\alpha\beta}g^{\beta\xi}
 \right)\w dx^{\rho\sigma}
 +\sqrt{-g}\left(\varepsilon_{\mu\nu\rho\sigma}g^{\nu\eta}\tilde{\varGamma}^\xi{}_\eta
 -\varepsilon_{\eta\nu\rho\sigma}g^{\nu\xi}\tilde{\varGamma}^\eta{}_\mu
 \right)\w dx^{\rho\sigma} \nonumber \\
 &\stackrel{\sigma}{=}0,
\label{G-term}
\end{align}
where we have used (\ref{dgGamma}) 
, substituted $\tilde{\varGamma}^{\mu}{}_\nu=\tilde{\varGamma}^{\mu}{}_{\nu\alpha}dx^\alpha$,
and then used the Hodge star relation, 
\begin{align}
 (\ast dx^\sigma):= \textstyle
 \frac{1}{3!}\sqrt{-g}g^{\sigma\tau}{\varepsilon}_{\tau\mu\nu\rho}, \quad 
 dx^{\mu\nu\rho}= \textstyle
 \frac{1}{\sqrt{-g}}{\varepsilon^{\mu\nu\rho\tau}} g_{\tau\sigma}
 (\ast dx^\sigma).
\label{Hodge}
\end{align}

Consequently, we obtain the Euler-Lagrange equations as,
\begin{align}
 0&=d\left\{
 \frac{1}{2\kappa}\varepsilon_{\mu\nu\rho\sigma}\sqrt{-g}
 \left(\tilde{R}^{\mu\nu} \w dx^\rho + \frac{2\lambda}{3!} dx^{\mu\nu\rho}\right)
 \right\} 
\label{Ein0} \\
 0&=-\frac12 \varepsilon_{\mu\nu\rho\sigma}g_{\xi\eta}\tilde{R}^{\mu\nu} \w dx^{\rho\sigma} 
 +\frac12 \left(
  \varepsilon_{\mu\eta\rho\sigma}\tilde{R}^\mu{}_\xi 
 +\varepsilon_{\mu\xi\rho\sigma}\tilde{R}^\mu{}_\eta 
 \right)\w dx^{\rho\sigma}
 +2
 g_{\xi\eta}\lambda dx^{0123}. 
\label{Ein1}
\end{align}
The pull back of these equations by $\sigma(x)$ 
are,
\begin{align}
 0&=d\left\{
 \frac{1}{\kappa}
 \left(
 G_{\sigma\xi} - \lambda g_{\sigma\xi}
 \right) (\ast dx^\xi) 
 \right\} 
, \quad G_{\sigma\xi}:= R_{\sigma \xi} - \frac{1}{2}g_{\sigma \xi}r,
\label{eq_Einstein0} \\
 0&=(r g_{\xi\eta}-2R_{\xi\eta}+2\lambda g_{\xi\eta})dx^{0123}
  =-2(G_{\xi\eta}-\lambda g_{\xi\eta})dx^{0123}, 
\label{eq_Einstein}
\end{align}
where $\ast$ is the Hodge operator (\ref{Hodge}).

The equation (\ref{eq_Einstein}) is the usual Einstein equation, 
and therefore, we may say that $(\ref{Ein1})$ is the Einstein equation with extended covariance.
By the discussions in the previous section, equation $(\ref{Ein0})$ 
coming from the variation with respect to $x^\mu$ 
should be considered as a conservation law of the energy-momentum current.
Let us define
by $\tilde{J}^G$, the energy-momentum current of the gravitational field
and by $\tilde{J}^\lambda$, that of the cosmological term, 
namely,
\begin{align}
 \tilde{J}^G_\sigma:=\frac1{2\kappa}\varepsilon_{\mu\nu\rho\sigma}\sqrt{-g}
  \tilde{R}^{\mu\nu} \w dx^\rho, \quad
  \tilde{J}^\lambda_\sigma
  :=\frac{1}{3!\kappa}\lambda \varepsilon_{\mu\nu\rho\sigma}
 \sqrt{-g}dx^{\mu\nu\rho}. 
\label{e-current1}
\end{align}
Then, the equation $(\ref{Ein0})$ says that the total energy-momentum current: 
$\tilde{J}_\sigma=\tilde{J}^G_\sigma+\tilde{J}^\lambda_\sigma$
satisfies the covariant energy-momentum conservation law, 
$0=d\tilde{J}_\sigma$.
To consider on the parameter space, namely, in the $x^\mu$ coordinates, 
take the pull back by $\sigma(x)$,
\begin{align}
 0=
d\left(J^G_\sigma+J^\lambda_\sigma\right),
 \quad
 J^G_\sigma:=\sigma^\ast \tilde{J}^G_\sigma
 =\frac{1}{\kappa}
 G_{\sigma\xi}(\ast dx^\xi), \quad
 J^\lambda_\sigma:=\sigma^\ast \tilde{J}^\lambda_\sigma
 =-\frac{\lambda}{\kappa} g_{\sigma\xi} (\ast dx^\xi). 
\label{e-hozon}
\end{align}
The above expression of energy-momentum of general relativity is one of the main results of 
the application of covariant Lagrangian formulation. 

Among these 
equations (\ref{eq_Einstein0}) and (\ref{eq_Einstein}), 
six equations are mutually independent, 
and the conventional view is to choose them 
from the 
Einstein equations (\ref{eq_Einstein}).
Actually,
when the Einstein equation (\ref{eq_Einstein}) holds, 
{\it the total energy-momentum current $J_\sigma$ is zero, 
and its conservation equation (\ref{eq_Einstein0}) is automatically satisfied}. 
Does this mean that the equation $(\ref{e-hozon})$ is a tautology? 
We claim this is not the case. 
Remember that the conservation law was obtained as a part of the Euler-Lagrange equations.
In the theory of extended covariance, there are no differences in their importance.

In such extended covariant perspective, 
Einstein's general relativity was just one case 
where a specific choice of parameterisation was made. 
The same goes for the choice of equation of motions. 
The equations $(\ref{Ein1})$ which corresponds to the 
balancing of stress energy-momentum tensor, were merely one choice for the fundamental equations, 
and there is no reason not to choose the others, $(\ref{Ein0})$. 
Actually, by using the relations 
$dg_{\alpha\beta}=g_{\xi\beta}{\varGamma^\xi}_\alpha+g_{\alpha\xi}{\varGamma^\xi}_\beta$, 
$d{R^\mu}_\nu+{\varGamma^\mu}_\xi \w {R^\xi}_\nu-{R^{\mu}}_\xi \w {\varGamma^\xi}_\nu=0$ 
and  
$dR^{\mu\nu}+{\varGamma^\mu}_\lambda \w R^{\lambda \nu}
+R^{\mu\lambda} \w {\varGamma^\nu}_{\lambda}=0$, 
the equation (\ref{eq_Einstein0}) becomes,
\begin{align}
 & d\left\{\varepsilon_{\mu\nu\rho\sigma}\sqrt{-g}\left(R^{\mu\nu} \w
 dx^\rho + \frac{2\lambda}{3!}dx^{\mu\nu\rho}\right)\right\} \nonumber\\
 &= \varepsilon_{\mu\nu\rho\sigma}
 \frac{\sqrt{-g}}{2}g^{\alpha\beta}dg_{\alpha\beta} \w 
 \left(
 R^{\mu\nu}\w dx^\rho+\frac{2\lambda}{3!}dx^{\mu\nu\rho}
 \right) 
 +\varepsilon_{\mu\nu\rho\sigma} \sqrt{-g}dR^{\mu\nu} \w dx^\rho,\nonumber\\
 &=
 -2\sqrt{-g}{\varGamma^{\mu\xi}}_\sigma \left(
{R}_{\mu\xi}-\frac12 r g_{\mu\xi}-\lambda g_{\mu\xi}
 \right) dx^{0123},  \label{deform}
\end{align}
which is just a linear combination of the standard Einstein equations,
and is equivalent to the four degrees of freedom (\ref{eq_Einstein}).

\subsubsection*{Gauge symmetry}

In our formulation, the general coordinate transformation is simply represented 
as a geometrical symmetry of Kawaguchi metric.
Let us consider a vector field:
\begin{align}
  {\cal G}=f^\mu \bib{}{x^\mu}+\left(\bib{f^\mu}{x^\rho}g^{\rho\nu}
 +\bib{f^\nu}{x^\rho}g^{\mu\rho}\right)\bib{}{g^{\mu\nu}},
\end{align}
where $f^\mu$ are functions of $x^\mu$.
This is a generator of the gauge transformation of the Kawaguchi metric.
We can show easily that $g_{\alpha\beta}dx^\alpha dx^\beta$ is {\it gauge invariant} under 
this transformation, {\it i.e.} ${\cal L}_{\cal G}(g_{\alpha\beta}dx^\alpha dx^\beta)=0$. 
We obtain the following transformation laws:
\begin{eqnarray*}
 &&{\cal L}_{\cal G} \tilde{\varGamma}^\mu{}_\xi
  =({\cal L}_{\cal G}g^{\mu\zeta}) \tilde{\varGamma}_{\zeta\xi\eta}dx^\eta
  +g^{\mu\zeta} ({\cal L}_{\cal G} \tilde{\varGamma}_{\zeta\xi\eta})dx^\eta
  +g^{\mu\zeta}\tilde{\varGamma}_{\zeta\xi\eta}d{\cal L}_{\cal G}x^\eta, 
 \\
 &&
 {\cal L}_{\cal G}\tilde{\varGamma}_{\zeta\xi\eta}
 \stackrel{\sigma}{=}
 -(\partial_\xi \partial_\eta f^\mu)g_{\zeta\mu}
 -(\partial_\zeta f^\mu)\tilde{\varGamma}_{\mu\xi\eta}
 -(\partial_\xi f^\mu)  \tilde{\varGamma}_{\zeta\mu\eta}
 -(\partial_\eta f^\mu) \tilde{\varGamma}_{\zeta\xi\mu},
 \\
 &&
 {\cal L}_{\cal G} \tilde{\varGamma}^\mu{}_\xi
 \stackrel{\sigma}{=}
 -\partial_\xi \partial_\eta f^\mu dx^\eta
 +(\partial_\zeta f^\mu)\tilde{\varGamma}^\zeta{}_{\xi}
 -(\partial_\xi f^\zeta)  \tilde{\varGamma}^\mu{}_{\zeta}, 
\end{eqnarray*}
Then, we can calculate the transformation of $\tilde{R}^{\mu\nu}$, 
\begin{eqnarray*}
 {\cal L}_{\cal G} \tilde{R}^\mu{}_\xi
 &=&d \left(
 {\cal L}_{\cal G}\tilde{\varGamma}^\mu{}_\xi
 \right) 
 +\left( {\cal L}_{\cal G}\tilde{\varGamma}^\mu{}_\lambda \right)
 \w \tilde{\varGamma}^\lambda{}_\xi
 +\tilde{\varGamma}^\mu{}_\lambda \w 
 \left( {\cal L}_{\cal G}\tilde{\varGamma}^\lambda{}_\xi \right)  \\
 &\stackrel{\sigma}{=}&
 (\partial_\zeta f^\mu)  \tilde{R}^\zeta{}_\xi
 -(\partial_\xi f^\zeta) \tilde{R}^\mu{}_\zeta
 \\
 {\cal L}_{\cal G} \tilde{R}^{\mu\nu}
 &\stackrel{\sigma}{=}&
  (\partial_\zeta f^\mu)  \tilde{R}^{\zeta\nu}
 +(\partial_\zeta f^\nu)  \tilde{R}^{\mu\zeta}.
\end{eqnarray*}
This is equivalent to the standard transformation law of the Riemann curvature.
Then, the condition ${\cal L}_{\cal G} K =0$ can be checked as follows, 
\begin{eqnarray*}
 {\cal L}_{\cal G} K &=&
 \left(
  {\cal L}_{\cal G}\frac{1}{4\kappa} 
  \sqrt{-g}\varepsilon_{\mu\nu\rho\sigma}dx^{\rho\sigma}
 \right) \w \tilde{R}^{\mu\nu}
 +\frac{1}{4\kappa} 
  \sqrt{-g}\varepsilon_{\mu\nu\rho\sigma}dx^{\rho\sigma}
 \w \left( {\cal L}_{\cal G} \tilde{R}^{\mu\nu} \right)
 \\
 &\stackrel{\sigma}{=}&
 \frac{1}{4\kappa} \sqrt{-g}
 \left\{
 \delta_{\mu\nu}^{\alpha\beta}
 (\partial_\zeta f^\zeta) \tilde{R}^{\mu\nu}{}_{\alpha\beta}
 -\delta_{\mu\nu\rho}^{\alpha\beta\zeta}
 (\partial_\zeta f^\rho) \tilde{R}^{\mu\nu}{}_{\alpha\beta} 
 -2 \delta_{\mu\nu}^{\alpha\beta}
 (\partial_\zeta f^\mu) \tilde{R}^{\zeta\nu}{}_{\alpha\beta}
 \right\} dx^{0123} \\
 &\stackrel{\sigma}{=}&
 \frac{\sqrt{-g}}{4\kappa} 
 \left\{
 2(\partial_\zeta f^\zeta) \tilde{R}^{\mu\nu}{}_{\mu\nu}
 -2(\partial_\zeta f^\zeta) \tilde{R}^{\mu\nu}{}_{\mu\nu}
 -4(\partial_\mu f^\rho) \tilde{R}^{\mu\nu}{}_{\nu\rho}
 -4(\partial_\zeta f^\mu) \tilde{R}^{\zeta\nu}{}_{\mu\nu}
 \right\} dx^{0123}\stackrel{\sigma}{=}0,
\end{eqnarray*}
where we have used
\begin{align}
& \tilde{R}^{\mu\nu}
 \stackrel{\sigma}{=}
 \frac12 \tilde{R}^{\mu\nu}{}_{\alpha\beta} dx^{\alpha\beta}, \\
& \tilde{R}^{\mu\nu}{}_{\alpha\beta} 
 :=
  \frac{\varepsilon_{\alpha\xi\eta\zeta}}{3!}
  \frac{dx^{\xi\eta\zeta} \w d\tilde{\varGamma}^{\mu\nu}{}_\beta}{dx^{0123}}
 -\frac{\varepsilon_{\beta\xi\eta\zeta}}{3!}
  \frac{dx^{\xi\eta\zeta} \w d\tilde{\varGamma}^{\mu\nu}{}_\alpha}{dx^{0123}}
 + \tilde{\varGamma}^{\mu}{}_{\lambda\alpha} \tilde{\varGamma}^{\lambda\nu}{}_\beta 
 - \tilde{\varGamma}^{\mu}{}_{\lambda\beta} \tilde{\varGamma}^{\lambda\nu}{}_\alpha.
\end{align}
It is easy to see that
the conservation law of the Noether current becomes,
\begin{align}
 &0=d\left[
 \frac{1}{2\kappa} \varepsilon_{\mu\nu\rho\sigma}\sqrt{-g}
 \left\{
 f^\sigma \left(\tilde{R}^{\mu\nu} \w dx^\rho+\frac{2\lambda}{3!} dx^{\mu\nu\rho}\right)
 -\frac12 \left(
 \bib{f^\mu}{x^\zeta}\tilde{\varGamma}^{\zeta\nu} 
 - g^{\nu\xi} \bib{f^\zeta}{x^\xi} \tilde{\varGamma}^\mu{}_\zeta
 \right) \w dx^{\rho\sigma}
 \right.\right. \nonumber \\
 &
 \left. \left.
 +\frac12 \frac{\partial^2 f^\mu}{\partial x^\xi \partial x^\eta}
 g^{\nu\xi}dx^{\eta\rho\sigma}
 \right\}
 \right].
\end{align}
Here, we used $\delta K={\cal L}_{\cal G}K=0$ and the covariant
Euler-Lagrange equations, $(\ref{Ein0})$ and $(\ref{Ein1})$.
This conservation law can be rewritten as,
\begin{align}
 &0=f^\sigma d\left\{
 \frac{1}{2\kappa} \varepsilon_{\mu\nu\rho\sigma}\sqrt{-g}
 \left(\tilde{R}^{\mu\nu} \w dx^\rho+\frac{2\lambda}{3!} dx^{\mu\nu\rho}\right) 
 \right\} \nonumber \\
 &
 -\bib{f^\chi}{x^\zeta}
 \left[
 \frac{1}{2\kappa} \varepsilon_{\mu\nu\rho\chi}\sqrt{-g}
 \left(\tilde{R}^{\mu\nu} \w dx^{\rho\zeta}+\frac{2\lambda}{3!} dx^{\mu\nu\rho\zeta}\right) 
 +\frac1{4\kappa} d \left\{
 \varepsilon_{\mu\nu\rho\sigma} \sqrt{-g}
 \left(
 \delta^\mu_\chi 
 \tilde{\varGamma}^{\zeta\nu}
 -g^{\nu\zeta}  \tilde{\varGamma}^{\mu}{}_\chi
 \right)\w dx^{\rho\sigma}
 \right\}
 \right] \nonumber \\
 &
 +\frac12 \frac{\partial^2 f^\chi}{\partial x^\zeta \partial x^\eta}
 \left[
 \frac1{2\kappa}\varepsilon_{\mu\nu\rho\sigma}\sqrt{-g}
 \left(\delta^\mu_\chi \tilde{\varGamma}^{\zeta\nu}
 -g^{\nu\zeta}\tilde{\varGamma}^\mu{}_\chi \right) \w dx^{\rho\sigma\eta}
 +\frac1{2\kappa}
 d\left(\varepsilon_{\mu\nu\rho\sigma} \sqrt{-g} 
\delta_\chi^\mu g^{\nu\zeta} dx^{\eta\rho\sigma}\right)
 \right].
\end{align}
Since $f^\sigma(x^\mu)$ are arbitrary functions of $x^\mu$,
$f^\sigma$ and its derivative terms must vanish separately.
This is the second Noether's theorem.
It means that, we can obtain the conservation law of 
the energy-momentum current 
$\tilde{J}_\sigma=\tilde{J}^G_\sigma+\tilde{J}^\lambda_\sigma$ 
also from the gauge symmetry (diffeomorphism invariance
of general relativity).
This is the similar mechanism when we derived the charge conservation law 
in the Maxwell-Dirac theory
from the $U(1)$ gauge symmetry. 

The gauge transformation of the energy-momentum current is,
\begin{align}
 {\cal L}_{\cal G} \tilde{J}^G_\sigma
 &=\frac{1}{2\kappa}\varepsilon_{\mu\nu\rho\sigma}
 \left\{
 ({\cal L}_{\cal G}\sqrt{-g})
 \tilde{R}^{\mu\nu}\w dx^\rho
 + \sqrt{-g} ({\cal L}_{\cal G}\tilde{R}^{\mu\nu}) \w dx^\rho
 + \sqrt{-g} \tilde{R}^{\mu\nu} \w ({\cal L}_{\cal G}dx^\rho)
 \right\} \nonumber \\
 &\stackrel{\sigma}{=}
 - \frac{1}{\kappa} (\partial_\sigma f^\rho) \tilde{G}_{\chi\rho} (*dx^\chi),\\
{\cal L}_{\cal G} \tilde{J}^\lambda_\sigma
 &=\frac{1}{3!\kappa}\lambda \varepsilon_{\mu\nu\rho\sigma}
\left\{ \left( {\cal L}_{\cal G}\sqrt{-g}  \right) 
dx^{\mu\nu\rho} +\sqrt{-g}{\cal L}_{\cal G}\left( dx^{\mu\nu\rho} \right)    \right\}
 \stackrel{\sigma}{=} \frac{\lambda}{\kappa} (\partial_\sigma f^\rho) g_{\rho\xi} (*dx^\xi),
\end{align}
where we set,
\begin{align}
\tilde{G}_{\xi\eta}
 &:=-\frac14 \varepsilon_{\mu\xi\rho\sigma}\tilde{R}^\mu{}_\eta \w dx^{\rho\sigma}
 -\frac14 \varepsilon_{\mu\eta\rho\sigma}\tilde{R}^\mu{}_\xi \w dx^{\rho\sigma}
 +\frac14 \varepsilon_{\mu\nu\rho\sigma}\tilde{R}^{\mu\nu} \w dx^{\rho\sigma}g_{\xi\eta}.
\end{align}
We obtain ${\cal L}_{\cal G} \tilde J_\sigma 
={\cal L}_{\cal G} (\tilde J^G_\sigma +\tilde{J}^\lambda_\sigma)$
$\stackrel{\sigma}{=}
 -\displaystyle{\frac{1}{\kappa}}(\partial_\sigma f^\rho) 
(\tilde{G}_{\xi\rho}-\lambda g_{\xi\rho}) (*dx^\xi)$.
The energy-momentum current is gauge invariant on the 
$4$-dimensional submanifold $\bm{\sigma}$ satisfying the 
equation of motion (\ref{eq_Einstein}).

\subsubsection*{Einstein-scalar field theory}

Here we will combine the Einstein's general relativity and the scalar field theory.
We will use the Kawaguchi metric:
\begin{align}
 K=\frac1{4\kappa} \varepsilon_{\mu\nu\rho\sigma}\sqrt{-g}~g^{\nu\alpha}
 \tilde{R}^{\mu}{}_\alpha \w dx^{\rho\sigma}
 -
{\frac{1}{\sqrt{-g}}}\frac{(d\phi \w dx_{\mu\nu\rho})(d\phi \w dx^{\mu\nu\rho})}
 {2\cdot 3! dx^{0123}}-V(\phi)\sqrt{-g}dx^{0123},
\end{align}
where we have defined 
$dx_{\mu\nu\rho}=g_{\mu\alpha}g_{\nu\beta}g_{\rho\gamma}dx^{\alpha\beta\gamma}$,
and 
the cosmological term is absorbed in the potential term $V(\phi)$.
As in the previous discussion, the variables $g^{\mu \nu}$ and $g_{\mu \nu}$ 
just represent coordinate functions and functions described by coordinate functions, respectively. 
The variation of $K$ now becomes, 
\begin{align}
\delta K &= \frac{\sqrt{-g}}{2}\left\{ -\frac{1}{\kappa}\tilde G_{\xi\eta} 
+ \tilde T_{\xi\eta}^{\phi}  \right\} \delta g^{\xi\eta}
+ d\left[ \frac{1}{4\kappa}\varepsilon_{\mu\nu\rho\sigma}\sqrt{-g}g^{\nu\xi}
\delta\tilde\Gamma^\mu{}_\xi \wedge dx^{\rho\sigma}   \right] \nonumber\\
&+ \delta\tilde\varGamma^\mu{}_\xi\w (\sigma^*{\rm \ vanishing\ term})
- d\left[( \tilde J_\xi^G +\tilde J_\xi^\phi   ) \delta x^\xi \right]
 + \left\{ d ( \tilde J_\xi^G +\tilde J_\xi^\phi   ) \right\} \delta x^\xi \nonumber\\
&- d\left[ \delta\phi {\frac{1}{\sqrt{-g}}} \frac{d\phi\wedge dx_{\mu\nu\rho}}{3!dx^{0123}} dx^{\mu\nu\rho}
  \right]
 + \delta\phi\left\{ d\left( {\frac{1}{\sqrt{-g}}} \frac{d\phi\wedge dx_{\mu\nu\rho}}{3!dx^{0123}} dx^{\mu\nu\rho} \right) -\sqrt{-g}V^\prime dx^{0123}
  \right\},  
\end{align}
where we set,
\begin{align} 
\tilde{T}_{\xi\eta}^{{\phi}}&:=
 \frac{(d\phi\w dx_{\xi\nu\rho})(d\phi \w {dx_\eta}^{\nu\rho})}{ {(-g)} 2! dx^{0123}}
 +\left\{
 {-} \frac{(d\phi \w dx_{\mu\nu\rho})(d\phi \w dx^{\mu\nu\rho})}{{(-g)}2\cdot 3!dx^{0123}}
 + V(\phi)
 \right\} g_{\xi\eta}, \\
\tilde{J}^\phi_\xi &:=
 -{\frac{1}{\sqrt{-g}}}\left\{ \frac{d\phi \w dx_{\mu\nu\xi}}{2! dx^{0123}}d\phi\wedge dx^{\mu\nu}
 +\left(
 -\frac{(d\phi \w dx_{\mu\nu\rho})(d\phi \w dx^{\mu\nu\rho})}{2 \cdot 3! (dx^{0123})^2}
 +V(\phi) 
 \right)\frac{\varepsilon_{\alpha\beta\gamma\xi}}{3!}dx^{\alpha\beta\gamma}\right\}.
\end{align}
The Euler-Lagrange equations are obtained as,
\begin{align}
 0&=-\frac1{\kappa}\tilde{G}_{\xi\eta}+\tilde{T}_{\xi\eta}^{{\phi}}, \\
 0&=-\sqrt{-g}V'dx^{0123}+d\left({\frac{1}{\sqrt{-g}}}\frac{d\phi\w dx_{\mu\nu\rho}}{3!dx^{0123}}
   dx^{\mu\nu\rho}\right), \\
 0&=d \left( \tilde{J}^G_\xi+\tilde{J}^\phi_\xi\right). 
\end{align}

\subsubsection*{Einstein-Maxwell field theory}

The Einstein-Maxwell field theory is described by,
\begin{align}
 &M=\{(x^\mu,g_{\mu\nu},A_\mu)\}=\{(z^a)\} \simeq \mathbb{R}^{18}, \quad
 K=K_{\rm Einstein}+ K_{\rm Maxwell}, \\ 
 &K_{\rm Einstein}=\frac1{4\kappa} \sqrt{-g} \varepsilon_{\mu\nu\rho\sigma}~g^{\nu\alpha}
 \tilde{R}^{\mu}{}_\alpha \w dx^{\rho\sigma}
 -\frac{\lambda}{\kappa} \sqrt{-g}~dx^{0123}, \\
 &
 K_{\rm Maxwell}
 = \frac{1}{\sqrt{-g}} \frac{(\tilde{F}\w dx_{\rho\sigma})
 (\tilde{F} \w dx^{\rho\sigma})}{4dx^{0123}}, \quad (dx^{0123}\neq 0),
\end{align}
where we have defined $dx_{\rho\sigma}=g_{\rho\alpha}g_{\sigma\beta}dx^{\alpha\beta}$. 
The Levi-Civita symbol is defined by $\varepsilon^{0123}=1$ and 
 $\varepsilon_{0123}=-1$. 
The variation of $K$ becomes, 
\begin{align}
\delta K &= \frac{\sqrt{-g}}{2}\left\{ -\frac{1}{\kappa}\tilde G_{\xi\eta}
 +\frac{1}{\kappa}\lambda g_{\xi\eta}dx^{0123} + \tilde T_{\xi\eta}^{\rm EM}  \right\} \delta g^{\xi\eta}
+ d\left[ \frac{1}{4\kappa} \sqrt{-g}\varepsilon_{\mu\nu\rho\sigma}g^{\nu\xi}\delta\tilde\Gamma^\mu{}_\xi \wedge dx^{\rho\sigma}   \right] \nonumber\\
&+ \delta\tilde\Gamma^\mu{}_\xi(\sigma^*{\rm \ vanishing\ term})
- d\left[( \tilde J_\sigma^G +\tilde J_\sigma^\lambda +\tilde J_\sigma^M+\tilde J_\sigma^A  ) \delta x^\sigma \right]
 + \left\{ d ( \tilde J_\sigma^G +\tilde J_\sigma^\lambda +\tilde J_\sigma^M+\tilde J_\sigma^A ) \right\} \delta x^\sigma \nonumber\\
&+ d\left[ \delta A_\mu \frac{1}{\sqrt{-g}} \frac{\tilde F \wedge dx_{\rho\sigma}}{2dx^{0123}} dx^{\rho\sigma\mu}
  \right]
 - \delta A_\mu  d\left( \frac{1}{\sqrt{-g}} \frac{\tilde F \wedge dx_{\rho\sigma}}{2dx^{0123}} 
 dx^{\rho\sigma\mu} \right),
\end{align}
where we set,
\begin{align}
 \tilde{J}^{M}_\sigma
 &:= \frac{1}{\sqrt{-g}}\left\{
  \frac{\tilde{F} \w dx_{\rho\sigma}}{dx^{0123}} \tilde{F} \w dx^\rho
 +\varepsilon_{\mu\nu\rho\sigma}
 \frac{(\tilde{F} \w dx_{\alpha\beta})(\tilde{F}\w dx^{\alpha\beta})}{4\cdot 3!(dx^{0123})^2}
 dx^{\mu\nu\rho}
 \right\}, 
 \\
 \tilde{J}^{A}_\sigma
 &:=
\frac{1}{\sqrt{-g}} 
 \frac{\tilde{F} \w dx_{\mu\nu}}{2 dx^{0123}} dx^{\mu\nu}\w dA_\sigma,
 \\
 \tilde{T}_{\xi\eta}^{\rm EM}
 &:= \frac{1}{g}
 \left\{
  \frac{(\tilde{F} \w dx_{\xi\sigma})(\tilde{F} \w {dx_{\eta}}^\sigma)}{dx^{0123}}
 -g_{\xi\eta}
 \frac{(\tilde{F} \w dx_{\rho\sigma})(\tilde{F}\w dx^{\rho\sigma})}{4dx^{0123}}
 \right\}.
\end{align}
The Euler-Lagrange equations become,
\begin{align}
 0&=d\left(\tilde{J}^G_\sigma+\tilde{J}^\lambda_\sigma+\tilde{J}^{M}_\sigma
 +\tilde{J}^{A}_\sigma\right), \\
 0&= -\frac{1}{\kappa}\tilde{G}_{\xi\eta}
 +\frac{1}{\kappa}\lambda g_{\xi\eta}dx^{0123}+\tilde{T}_{\xi\eta}^{\rm EM}, \\
 0&=d\left(\frac{1}{\sqrt{-g}}
 \frac{\tilde{F} \w dx_{\rho\sigma}}{2dx^{0123}}dx^{\rho\sigma\mu}\right).
\label{EOM}
\end{align}
Since $dx^{0123}\neq 0$ is assumed for $K_{\rm Maxwell}$, 
$\displaystyle{\frac{\partial (x^0,x^1,x^2,x^3)}{\partial (s^0,s^1,s^2,s^3)}\neq 0}$ 
is satisfied for any parameterisation $\sigma(s)$.
Therefore, the matrix $\displaystyle{\left(  \frac{\partial x^\mu}{\partial s^\alpha} \right)}$ 
is invertible.
Thus,
from (\ref{EOM}),
\begin{eqnarray*}
 0&=&
\sigma^\ast(s) 
 d\left(\frac{1}{\sqrt{-g}}\frac{\tilde{F} \w dx_{\rho\sigma}}{2dx^{0123}}dx^{\rho\sigma}\right)
 \left( \frac{\partial x^\mu}{\partial s^\alpha}  \right) \w ds^\alpha  
 ~\Rightarrow~   
 0 = 
\sigma^\ast(s)
 d\left(\frac{1}{\sqrt{-g}}\frac{\tilde{F} \w 
 dx_{\rho\sigma}}{2dx^{0123}}dx^{\rho\sigma}\right),
\end{eqnarray*}
and if $\bm{\sigma}$ is the $4$-dimensional submanifold satisfying the Euler-Lagrange equations, 
we get,
\begin{align}
 \sigma^\ast \tilde{J}^{A}_\sigma=
 d \left\{
 \sigma^\ast
 \left(
 \frac{1}{\sqrt{-g}} \frac{\tilde{F} \w dx_{\mu\nu}}{2dx^{0123}} dx^{\mu\nu} A_\sigma
 \right)\right\},
\end{align}
which is the exact form.
We obtain two conservation laws:
\begin{align}
 d \left( \sigma^\ast \tilde{J}^{A}_\sigma\right) & = 0, \nonumber \\
 d \left( \sigma^\ast \tilde{J}^G_\sigma
 +\sigma^\ast \tilde{J}^\lambda_\sigma
 +\sigma^\ast\tilde{J}^{M}_\sigma \right) 
 &= 0.
\end{align}
However, this happens due to the specific form of  
$K_{\rm Maxwell}$, and for more general model such as Born-Infeld, 
such separation of conserved currents does not occur.

\subsubsection*{Einstein coupled to perfect fluid} 

Following the standard theory such as~\cite{Lanczos, Choquet}, 
we construct the extended configuration space for the perfect fluid 
under gravitational force as $M=\{(x^\mu,g^{\mu\nu},\pi_\mu,p)\}$ with the Kawaguchi metric 
\begin{align}
 &K=K_{\rm Einstein}+K_{\rm P.F},\quad
 K_{\rm P.F.}=\left(\rho+p\right) \sqrt{-g} dx^{0123}+ \int \delta K_{p},  \label{E-PF-K}
 \\ 
 &
 \delta K_p:=
 p\delta \left(
 \sqrt{-g}dx^{0123}
 \right)
 + p d\left(-\frac{1}{3!}\varepsilon_{\mu\nu\rho\sigma}\sqrt{-g}dx^{\mu\nu\rho}\right)
\delta x^\sigma, \label{E-PF-dK}
\end{align}
where $K_{\rm Einstein}$ is given by (\ref{KEmetric}) 
and the integral in (\ref{E-PF-K}) denotes a formal functional integral such that gives (\ref{E-PF-dK}) 
when taken the variation. 
The $\rho$ and $p$ are the usual energy density and pressure density, 
but since we prefer to work with the variable $\pi_\mu$ following ~\cite{Dirac}, 
$\rho$ is rather defined by this co-vector field $\pi_\mu$ by 
\begin{align}
 \left(\rho+p\right)\sqrt{-g}:=\sqrt{g^{\mu\nu}\pi_\mu \pi_\nu}. 
\end{align}
The commonly used $4$-velocity field $u$ is related to $\pi_\mu$ by 
$\pi_\mu=:\left(\rho+p\right) \sqrt{-g}u_\mu.$ 
By this definition, we automatically get $g^{\mu\nu}u_\mu u_\nu=1$. 

The variation of $\pi_\mu$ is defined as the following to satisfy the generalised 
Lin constraints~\cite{Marsden}, 
\begin{align}
 \delta \pi_\mu
 =
 \pi^\alpha \bib{\delta \xi_\mu}{x^\alpha}
 -{\tilde{\varGamma}{}^\alpha}_{\mu\beta}\delta \xi_\alpha \pi^\beta
 +\sqrt{-g}g^{\alpha\beta}\delta \xi_\beta\bib{p}{x^\alpha} u_\mu,
 \quad \pi^\mu:=g^{\mu\nu}\pi_\nu, \label{lin}
\end{align}
Here the vectors $\displaystyle{\delta \xi:=\delta \xi^\mu \bib{}{x^\mu}}, 
\,\delta \xi^\mu:=g^{\mu\nu} \delta \xi_\nu$ represents the infinitesimal displacement of the 
fluid element, which satisfies the relation:
\begin{align}
 \delta \pi_\mu = -({\cal L}_{\delta \xi}\pi)_\mu, \quad \pi:=\pi_\mu dx^\mu. 
\end{align}
We can now calculate the variation of $K_{P.F.}$ in terms of this $\delta \pi_\mu$, 
together with $\delta x^\mu$ and $\delta g^{\mu \nu}$,
\begin{align}
 \delta K_{\rm P.F.}&=
 \bib{}{x^\alpha}
 \left\{
 \left(\rho+p\right)
 \sqrt{-g}u^\alpha u^\mu
 \delta \xi_\mu
 \right\}dx^{0123}
 \nonumber \\
 &
 +\delta \xi_\mu \left[
 g^{\mu\nu}
 \bib{p}{x^\nu}
 -
 \frac{1}{\sqrt{-g}}
 \bib{}{x^\alpha}
 \left\{
 \sqrt{-g}
 \left(
 \rho+p
 \right) u^\alpha u^\mu
 \right\}
 -
 \left(\rho+p\right)
 {\tilde{\varGamma}{}^\mu}_{\alpha\beta}u^\alpha u^\beta
 \right]\sqrt{-g}dx^{0123}
 \nonumber \\ 
 &
 +
 \frac{\delta g^{\mu\nu}}2 \left\{
 \left(\rho+p\right) u_\mu  u_\nu
 -p g_{\mu\nu}
 \right\}\sqrt{-g}dx^{0123}
 \nonumber \\
 &
 +d
 \left\{
 \frac{1}{3!}
 \left(\rho+p\right)\sqrt{-g}
 \varepsilon_{\mu\nu\rho\sigma}
  dx^{\mu\nu\rho}
 \delta x^\sigma
 +\frac{1}{3!}p\sqrt{-g}\varepsilon_{\mu\nu\rho\sigma}dx^{\mu\nu\rho}
 \delta x^\sigma
 \right\}
 \nonumber \\
 &
 -
 \left[
 dp \w \left(-\frac{1}{3!}
 \sqrt{-g}
 \varepsilon_{\mu\nu\rho\sigma}
  dx^{\mu\nu\rho}\right)
 -
 d\left\{
 -\frac{1}{3!}\left(\rho+3p\right)
 \sqrt{-g}
 \varepsilon_{\mu\nu\rho\sigma}
  dx^{\mu\nu\rho}
 \right\}
 \right] \delta x^\sigma.
\end{align}
The Euler-Lagrangian equations pulled back by $\sigma(x)$ is then, 
\begin{align}
 dp\w J^g_\sigma
 &=
 d(J^G_\sigma+J^\lambda_\sigma+J^{P.F.}_\sigma), \label{EL1line}
 \\
 0&=-\frac{1}{\kappa}\left(G_{\mu\nu}-\lambda g_{\mu\nu}\right)+T^{P.F.}_{\mu\nu}, \label{EL2line}
 \\
 g^{\mu\nu}\bib{p}{x^\nu}
 &=\frac{1}{\sqrt{-g}}
 \bib{}{x^\alpha}\left\{\sqrt{-g}\left(\rho+p\right)u^\alpha u^\mu\right\}
 +{\varGamma^\mu}_{\alpha\beta}\left(\rho+p\right)u^\alpha u^\beta, \label{EL3line} 
\end{align}
where we set 
\begin{align}
 &J^{P.F.}_\sigma:=-\frac{1}{3!}\varepsilon_{\mu\nu\rho\sigma}
 \left(\rho+3p\right) \sqrt{-g}dx^{\mu\nu\rho}, \quad
 J^g_\sigma:=-\frac{1}{3!}\varepsilon_{\mu\nu\rho\sigma}
 \sqrt{-g}dx^{\mu\nu\rho},
 \\
 &T^{P.F.}_{\mu\nu}:=\left(\rho+p\right)u_\mu u_\nu-p g_{\mu\nu}.
\end{align}
The equation (\ref{EL3line}) can be rewritten as 
\begin{align}
 g^{\mu\nu}\bib{p}{x^\nu}&=
 \frac{1}{\sqrt{-g}}
 \partial_\alpha\left\{
 \sqrt{-g}
 \left(\rho+p\right)u^\alpha u^\mu\right\}
 +{\varGamma^\mu}_{\alpha\beta}
 \left(\rho+p\right)u^\alpha u^\beta
 \\
 &={\varGamma^\lambda}_{\lambda\alpha}\left(\rho+p\right)u^\alpha u^\mu
 +
 \partial_\alpha \left\{\left(\rho+p\right)u^\alpha \right\}u^\mu
 +\left(\rho+p\right)
 \left(
 u^\alpha \partial_\alpha u^\mu+{\varGamma^\mu}_{\alpha\beta}
 u^\alpha u^\beta
 \right)
 \\
 &=
 \nabla_\alpha \left\{\left(\rho+p\right)u^\alpha\right\}u^\mu
 +\left(\rho+p\right)\nabla_u u^\mu,
\end{align}
and contracting with $u_\mu$, we get 
\begin{align}
 u^\alpha \bib{p}{x^\alpha}&=
 \nabla_\alpha \left\{\left(\rho+p\right)u^\alpha\right\},
\end{align} 
then, substituting back, we get the relativistic Euler equation for the perfect fluid, 
\begin{align}
 \left(\rho+p\right)\nabla_u u^\mu=g^{\mu\nu}\bib{p}{x^\nu}
 -u^\nu \bib{p}{x^\nu}u^\mu.
\end{align}
Actually, it is easy to check the pull back Euler-Lagrange equation (\ref{EL3line}) 
is equivalent to $\nabla_\mu (T^{P.F.})^{\mu\nu}=0$.

In our framework, the energy-momentum relation (\ref{EL1line}) appears as a part of
the Euler-Lagrange equations. Here, $J^{P.F.}_\sigma$ stands for the energy-momentum current
of the fluid and the left hand side $dp\w J^g_\sigma$ expresses its dissipation.Take the differentiation of $J^{P.F.}_\sigma$,
\begin{align}
 dJ^{P.F.}_\sigma&=-\frac{1}{3!}\varepsilon_{\mu\nu\rho\sigma}\bib{}{x^\alpha}
 \left\{\left(\rho+3p \right)\sqrt{-g}\right\}dx^{\alpha\mu\nu\rho}
 =-\bib{}{x^\sigma}\left\{\left(\rho+3p\right)\sqrt{-g}\right\}dx^{0123} \nonumber
 \\
 &=-\bib{\sqrt{g^{\mu\nu}\pi_\mu \pi_\nu}}{x^\sigma}dx^{0123}
 -2\bib{(p\sqrt{-g})}{x^\sigma}dx^{0123} \nonumber
 \\
 &=\left\{-\frac{\pi^\mu \partial_\sigma \pi_\mu
 +\frac{1}{2}\partial_\sigma g^{\mu\nu} \pi_\mu \pi_\nu}
 {\sqrt{g^{\alpha\beta}\pi_\alpha \pi_\beta}}
  -2\bib{(p\sqrt{-g})}{x^\sigma}\right\}dx^{0123}, \label{dJPF}
\end{align}
then applying the Lin constraints (\ref{lin}) with $\delta \xi^\mu=\delta^\mu_\sigma$,
we can rewrite the first term in (\ref{dJPF}),
\begin{align}
 \partial_\sigma \pi_\mu&=-\pi^\alpha (\partial_\alpha g_{\mu\sigma})
 +{\varGamma}_{\sigma\mu\beta}\pi^\beta-\sqrt{-g}\bib{p}{x^\sigma}u_\mu \nonumber
 \\
 &=
 -\pi^\alpha ({\varGamma}_{\mu\sigma\alpha}+{\varGamma}_{\sigma\mu\alpha})
 +{\varGamma}_{\sigma\mu\beta}\pi^\beta-\sqrt{-g}\bib{p}{x^\sigma}u_\mu
 =-{\varGamma}_{\mu\sigma\alpha}\pi^\alpha-\sqrt{-g}\bib{p}{x^\sigma}u_\mu.
\end{align}
The equation (\ref{dJPF}) becomes
\begin{align}
dJ^{P.F.}_\sigma&=
\left\{
 2\sqrt{-g}{\varGamma^{\mu\nu}}_\sigma \left(\rho+p\right) u_\mu u_\nu
 +\sqrt{-g}\bib{p}{x^\sigma}
 -2\bib{(\sqrt{-g}p)}{x^\sigma}
 \right\}dx^{0123},
\end{align}
and together with $\displaystyle{dp \w J^g_\sigma=-\bib{p}{x^\sigma}\sqrt{-g}dx^{0123}}$, 
\begin{align}
-dp\w J^g_\sigma+dJ^f_\sigma
 &=2\sqrt{-g}{\varGamma^{\mu\nu}}_\sigma 
 \left\{(\rho+p)u_\mu u_\nu-pg_{\mu\nu}\right\}dx^{0123}
 =
 2\sqrt{-g}{\varGamma^{\mu\nu}}_\sigma
 T^{P.F.}_{\mu\nu}dx^{0123}.
\end{align}
Finally, the Euler-Lagrange equation (\ref{EL1line}) turns out to be
\begin{align}
 0=-dp \w J^g_\sigma+d(J^G_\sigma+J^\lambda_\sigma+J^{P.F.}_\sigma)
  =-2\sqrt{-g}{\varGamma^{\mu\nu}}_\sigma
 \left\{\frac{1}{\kappa}\left(R_{\mu\nu}-\frac12r g_{\mu\nu}-\lambda g_{\mu\nu}\right)-T^{P.F.}_{\mu\nu}\right\}
 dx^{0123},
\end{align}
which is a generalisation of (\ref{deform}).

\subsubsection*{Energy density of Schwarzschild spacetime} 

We will give an example of Schwarzchild spacetime for the energy current 
(\ref{e-current1}). 
In local coordinate system $(x^0, x^1, x^2, x^3)=(t,r,\th, \vp)$,
the Schwarzschild metric is given by, 
\begin{align}
 ds^2
 =\left(1-\fr{r_s}{r}\right)c^2dt^2-\left(1-\fr{r_s}{r}\right)^{-1}dr^2
 -r^2d\th^2-r^2\sin^2\th d\vp^2, 
\end{align}
where $r_s=2GM/c^2$ is the Schwarzschild radius.
Since there is a singularity at $r=0$, 
we used the distributional techniques to regularize 
$1/r$ following~\cite{Balasin-Nachbagauer, Kawai-Sakane},
\begin{align}
 \frac{1}{r}:=\lim_{\epsilon \to 0} \frac{1}{(r^2+\epsilon^2)^{\frac{1}{2}}},
 \quad
 \left(\frac1{r}\right)'=\lim_{\epsilon \to 0} \frac{-r}{(r^2+\epsilon^2)^{\frac32}}.
\end{align}
Using the expression (\ref{e-current1}), 
\begin{align}
 & J_0^G=-\frac{1}{2\kappa}\frac{2r_s \epsilon^2}{r^2(r^2+\epsilon^2)^\frac32}r^2 \sin{\theta}dr
 \w d\theta \w d\varphi,
\end{align}
and taking the limit $\varepsilon \to 0$, we get,
\begin{align}
 & J^G_0=-\sqrt{-g}Mc^2 \delta^{(3)}(\bm{x}) dx^{123}, \label{J0G} 
\end{align}
where we have used the following formulae~\cite{Kawai-Sakane},
\begin{eqnarray}
 \frac{r_s\epsilon^2}{r^2(r^2+\epsilon^2)^\frac32} 
 \xrightarrow[\epsilon \to 0]{} 4\pi r_s \delta^{(3)}(\bm{x}).
\end{eqnarray}
The energy density (\ref{J0G}) takes negative value,
which is exactly cancelled out by the matter energy density,
$J_0^{\rm matter}=\sqrt{-g}Mc^2\delta^{(3)}(\bm{x})dx^{123}$,
so that 
$J_0^G+J_0^{\rm matter}=0$ on-shell. 
This vanishing energy density comes from our specific but gauge invariant definition 
of energy-momentum current of gravity, given in (\ref{e-current1}) and (\ref{e-hozon}).

Here we mention a relation of our energy-momentum currents of gravity (\ref{e-current1})
to those by Dirac~\cite{Dirac}, where the 
calculation is carried out 
by
first order formulation. 
For comparison, we also calculate by first order in the Kawaguchi setting. 

The first order Kawaguchi metric of general relativity is given by
\begin{align}
 K_{\rm 1st.}=
 -\frac{1}{4\kappa} \varepsilon_{\mu\nu\rho\sigma}\sqrt{-g}
 g^{\nu\xi}\tilde{\varGamma}^\mu{}_\lambda \w \tilde{\varGamma}^\lambda{}_\xi
 \w dx^{\rho\sigma},   \label{1st-schwarz} 
\end{align}
and from $K_{\rm 1st.}$, the energy-momentum current is derived as 
\begin{align}
 \tilde{J}^{\rm 1st.}_\sigma=d\left(
 \frac{1}{2\kappa}\varepsilon_{\mu\nu\rho\sigma}\sqrt{-g}
 g^{\nu\xi}dx^{\rho}
 \right) \w \tilde{\varGamma}^\mu{}_\xi
 +
 \frac{1}{2\kappa}\varepsilon_{\mu\nu\rho\sigma}\sqrt{-g}g^{\nu\xi}
 \tilde{\varGamma}^\mu{}_\lambda \w \tilde{\varGamma}^\lambda{}_\xi \w dx^\rho. \label{J_1st}
\end{align}
The derivation details are given in Appendix B.
The energy current similarly calculated by distributional technique becomes,
\begin{align}
 & J_0^{\rm 1st.}=\frac{1}{2\kappa}\left\{
 -2+\frac{2r_s \epsilon^2}{(r^2+\epsilon^2)^\frac32}\right\}
 \sin{\theta}dr\w d\theta \w d\varphi \,
 \stackrel{\epsilon \to 0}{\longrightarrow} \, -\frac{1}{\kappa r^2}dx^{123} +Mc^2
 \sqrt{-g}\delta^{(3)}(\bm{x})dx^{123}. 
\end{align}
Taking the integral over the $t=0$ hypersurface, the first term gives a trivial
divergence and is neglected.
The second term gives a positive value for energy, $Mc^2$.
However, since the surface term is dropped in (\ref{1st-schwarz}), 
this expression is essentially coordinate dependent. 
On the other hand, our energy current of gravity (\ref{e-current1}) is 
{\it gauge invariant}
and {\it reparameterisation invariant}. 
Thus, it has a physical 
meaning ({\it i.e.} observable) of spacetime.

\subsubsection*{Palatini $f(R)$ gravity}

In general relativity, we found that the energy-momentum current is zero 
under the on-shell conditions. 
This property holds even in other theories of gravity such as Palatini $f(R)$ gravity:
\begin{align}
 L=-\fr{\sqrt{-g}}{2\kappa}f(R), \quad
 R=g^{\nu\rho}{R^\mu}_{\nu\mu\rho}, \quad
 {R^\lambda}_\mu=\frac12 {R^\lambda}_{\mu\nu\rho}dx^{\nu\rho}
 =d{\varGamma^\lambda}_\mu+{\varGamma^\lambda}_\xi \w {\varGamma^\xi}_\mu,
\end{align}
where the metric $g_{\mu\nu}$ and the linear connection ${\varGamma^\lambda}_\mu$
are independent. 
Namely, the metricity condition (\ref{connection}) does not hold. 
We will assume the torsion-free connection and
keep the last two indices of ${\varGamma}^{\lambda}{}_{\mu\nu}$ symmetric
as in \cite{Sotiriou}.
The Kawaguchi metric is given by, 
\begin{align}
 K=-\fr{\sqrt{-g}}{2\kappa} f
  \left(
  -\fr{\varepsilon_{\mu\nu\rho\sigma} \tilde{R}^{\mu\nu} \w dx^{\rho\sigma}}{2dx^{0123}}
  \right)
 dx^{0123}.
\end{align}
We obtain the variation of $K$ as follows,
\begin{align}
 \delta K
 &
 =\frac1{4\kappa} \sqrt{-g}
  \left(
  f g_{\xi\eta}dx^{0123}
  +f' \varepsilon_{\mu(\xi|\rho\sigma}\tilde{R}^\mu{}_{|\eta)} \w dx^{\rho\sigma}
  \right)
 \delta g^{\xi\eta}
 \nonumber \\
 &
 +d
  \left(
  \frac1{4\kappa} f' \varepsilon_{\mu\nu\rho\sigma}\sqrt{-g}g^{\nu\xi}
  \delta \tilde{\varGamma}^\mu{}_\xi \w dx^{\rho\sigma}
  \right)
 \nonumber \\
 &
 -\fr{1}{4\kappa} \delta \tilde{\varGamma}^\mu{}_{\lambda\tau}
  \biggl[
  d
   \left\{
   \sqrt{-g} f' \varepsilon_{\mu\nu\rho\sigma} g^{\nu(\lambda}dx^{\tau)\rho\sigma}
   \right\}
 \nonumber \\
 &
 \hspace{25mm}
  +\sqrt{-g} f'
   \left\{
   \varepsilon_{\mu\nu\rho\sigma} g^{\nu\xi} 
  \tilde{\varGamma}^{(\lambda}{}_{\xi} \w dx^{\tau)\rho\sigma}
   - \varepsilon_{\xi\nu\rho\sigma}g^{\nu (\lambda}
  \tilde{\varGamma}^{|\xi|}{}_{\mu} \w dx^{\tau)\rho\sigma}
   \right\}
  \biggr]
 \nonumber \\
 &
 -d
  \Biggl[
  \frac{1}{2\kappa}
   \biggl[
   \sqrt{-g} \varepsilon_{\mu\nu\rho\sigma}
    \left\{
    f' \tilde{R}^{\mu\nu} \w dx^\rho 
    +\fr{1}{3!}
     \left(
     f
     +f'\fr{\varepsilon_{\alpha\beta\gamma\delta} 
 \tilde{R}^{\alpha\beta} \w dx^{\gamma\delta}}{2dx^{0123}}
     \right)
    dx^{\mu\nu\rho}
    \right\}
 \nonumber \\
 &
 \hspace{20mm}
   -\fr12 \tilde{\varGamma}^{\mu}{}_{\lambda\sigma}
    \biggl\{
    d
     \left(
     \sqrt{-g} g^{\nu\lambda} f' \varepsilon_{\mu\nu\rho\tau} dx^{\rho\tau}
     \right)
 \nonumber \\
 &
 \hspace{40mm}
    +\sqrt{-g} f'
     \left(
     g^{\nu\xi} \tilde{\varGamma}^{\lambda}{}_{\xi\eta} \varepsilon_{\mu\nu\rho\tau}
     -g^{\nu\lambda} \tilde{\varGamma}^{\xi}{}_{\mu\eta} \varepsilon_{\xi\nu\rho\tau}
     \right)
    dx^{\eta\rho\tau}
    \biggr\}
   \biggr]
   \delta x^\sigma 
  \Biggr]
 \nonumber \\
 &
 +d
  \Biggl[
  \frac{1}{2\kappa}
   \biggl[
   \sqrt{-g} \varepsilon_{\mu\nu\rho\sigma}
    \left\{
    f' \tilde{R}^{\mu\nu} \w dx^\rho 
    +\fr{1}{3!}
     \left(
     f
     +f'\fr{\varepsilon_{\alpha\beta\gamma\delta} 
  \tilde{R}^{\alpha\beta} \w dx^{\gamma\delta}}{2dx^{0123}}
     \right)
    dx^{\mu\nu\rho}
    \right\}
 \nonumber \\
 &
 \hspace{20mm}
   -\fr12 \tilde{\varGamma}^{\mu}{}_{\lambda\sigma}
    \biggl\{
    d
     \left(
     \sqrt{-g} g^{\nu\lambda} f' \varepsilon_{\mu\nu\rho\tau} dx^{\rho\tau}
     \right)
 \nonumber \\
 &
 \hspace{40mm}
    +\sqrt{-g} f'
     \left(
     g^{\nu\xi} \tilde{\varGamma}^{\lambda}{}_{\xi\eta} \varepsilon_{\mu\nu\rho\tau}
     -g^{\nu\lambda} \tilde{\varGamma}^{\xi}{}_{\mu\eta} \varepsilon_{\xi\nu\rho\tau}
     \right)
    dx^{\eta\rho\tau}
    \biggr\}
   \biggr]
  \Biggr]
 \delta x^\sigma,
\end{align}
where the round brackets denotes symmetrisation: $A^{(\mu\nu)}=\fr12 (A^{\mu\nu}+A^{\nu\mu})$. 
The Euler-Lagrange equations become, 
\begin{align}
 0&=
 f g_{\xi\eta}dx^{0123}
 +f' \varepsilon_{\mu(\xi|\rho\sigma}\tilde{R}^\mu{}_{|\eta)} \w dx^{\rho\sigma}
 \label{fr1}
 \\
 0&=
 d
  \left\{
  \sqrt{-g} f' \varepsilon_{\mu\nu\rho\sigma} g^{\nu(\lambda}dx^{\tau)\rho\sigma}
  \right\}
 \nonumber \\
 &
 \hspace{5mm}
 +\sqrt{-g} f'
  \left\{
  \varepsilon_{\mu\nu\rho\sigma} 
  g^{\nu\xi} \tilde{\varGamma}^{(\lambda}{}_{\xi} 
  \w dx^{\tau)\rho\sigma}
  -\varepsilon_{\xi\nu\rho\sigma} g^{\nu (\lambda}
  \tilde{\varGamma}^{|\xi|}{}_{\mu} \w  
  dx^{|\tau)\rho\sigma}
  \right\}
 \label{fr2}
 \\
 0&=
 d\tilde{J}^{f(R)}_\sigma,
\end{align}
where 
$\tilde{J}^{f(R)}_\sigma$ is 
the energy-momentum current of the Palatini $f(R)$ gravity
defined by,
\begin{align}
 \tilde{J}^{f(R)}_\sigma
 &=
  \frac{1}{2\kappa}
   \biggl[
   \sqrt{-g} \varepsilon_{\mu\nu\rho\sigma}
    \left\{
    f' \tilde{R}^{\mu\nu} \w dx^\rho 
    +\fr{1}{3!}
     \left(
     f
     +f'\fr{\varepsilon_{\alpha\beta\gamma\delta} 
 \tilde{R}^{\alpha\beta} \w dx^{\gamma\delta}}{2dx^{0123}}
     \right)
    dx^{\mu\nu\rho}
    \right\}
 \nonumber \\
 &
 \hspace{20mm}
   -\fr12 \tilde{\varGamma}^{\mu}{}_{\lambda\sigma}
    \biggl\{
    d
     \left(
     \sqrt{-g} g^{\nu\lambda} f' \varepsilon_{\mu\nu\rho\tau} dx^{\rho\tau}
     \right)
 \nonumber \\
 &
 \hspace{40mm}
    +\sqrt{-g} f'
     \left(
     g^{\nu\xi} \tilde{\varGamma}^{\lambda}{}_{\xi\eta} \varepsilon_{\mu\nu\rho\tau}
     -g^{\nu\lambda} \tilde{\varGamma}^{\xi}{}_{\mu\eta} \varepsilon_{\xi\nu\rho\tau}
     \right)
    dx^{\eta\rho\tau}
    \biggr\}
   \biggr].
 \label{fr_current}
\end{align}
We recover the standard equations from the pull back of (\ref{fr1}) and (\ref{fr2}):
\begin{align}
 0&=
 \left[
 f'(R)R_{(\xi\eta)}-\fr12 f(R)g_{\xi\eta}
 \right]
 dx^{0123},
 \label{pfr1}
 \\
 0&=
  \left[
  \nabla_\mu
   \left\{
   \sqrt{-g} f' g^{\tau\lambda}
   \right\}
  -\nabla_\nu
   \left\{
   \sqrt{-g} f' g^{\nu (\lambda}
   \right\}
  \delta^{\tau )}_\mu
  \right]
 dx^{0123}.
 \label{pfr2}
\end{align}
The pull back of (\ref{fr_current}) becomes,
\begin{align}
 J^{f(R)}_\sigma
 &
 =\fr{1}{2\kappa}
  \biggl[f'(R)\left\{
  R_{\xi\sigma}-R g_{\xi\sigma}-R^{\mu\nu}{}_{\nu\sigma} g_{\mu\xi}\right\}
  -g_{\sigma\xi}
   \left\{
   f(R)-f'(R)R
   \right\}
 \nonumber \\
 &
 \hspace{10mm}
  +\fr{1}{\sqrt{-g}} \varGamma^\mu {}_{\lambda\sigma}
   \left\{
   \nabla_\mu
    \left(
    \sqrt{-g} g^{\nu\lambda} f'(R)
    \right)
   g_{\nu\xi}
   -
   \nabla_\nu
    \left(
    \sqrt{-g} g^{\nu\lambda} f'(R)
    \right)
   g_{\mu\xi}
   \right\}
  \biggr]
 (*dx^\xi).
 \label{pfrc}
\end{align}
As shown in ~\cite{Sotiriou}, equation (\ref{pfr2}) can be reduced to
\begin{align}
 \nabla_\mu
   \left\{
   \sqrt{-g} f' g^{\tau\lambda}
   \right\}=0.
\end{align}
It is equivalent to 
\begin{align}
 \nabla_\mu
   \left\{
   \sqrt{-h} h^{\tau\lambda}
   \right\}=0, \label{hmetric}
\end{align}
with new metric $h_{\mu\nu}:=f'(R)g_{\mu\nu}$.
The equation (\ref{hmetric}) is the metricity condition of $h_{\mu\nu}$, 
and it is solved as,
\begin{align}
 {\varGamma}^{\lambda}{}_{\mu\nu}
 =\fr12 h^{\lambda \sigma}
 (\partial_\mu h_{\nu \sigma}+\partial_\nu h_{\mu \sigma}-\partial_\sigma h_{\mu \nu}).
\end{align}
Therefore, we have the same symmetries for the curvature tensor
as general relativity; $R_{\mu\nu\rho\sigma}=-R_{\nu\mu\rho\sigma}$, etc.
Consequently, (\ref{pfrc}) becomes
\begin{align}
 J^{f(R)}_\sigma
 &
 =\fr{1}{2\kappa}
  \biggl[2f'(R)R_{\xi\sigma}-f(R)g_{\xi\sigma}
 \nonumber \\
 &
 \hspace{10mm}
  +\fr{1}{\sqrt{-g}} \varGamma^\mu {}_{\lambda\sigma}
   \left\{
   \nabla_\mu
    \left(
    \sqrt{-h} h^{\nu\lambda}
    \right)
   g_{\nu\xi}
   -
   \nabla_\nu
    \left(
    \sqrt{-h} h^{\nu\lambda}
    \right)
   g_{\mu\xi}
   \right\}
  \biggr]
 (*dx^\xi)=0, \label{palacurr}
\end{align}
by the use of (\ref{pfr1}) and metricity derived from (\ref{pfr2}).
The generator of gauge transformations in this case is given by, 
\begin{align}
 {\cal G}= & f^\mu
 \bib{}{x^\mu}+
 \left(\bib{f^\mu}{x^\rho}g^{\rho\nu}
 +\bib{f^\nu}{x^\rho}g^{\mu\rho}\right)
 \bib{}{g^{\mu\nu}} \nonumber
 \\
 &
 +\left(
-\frac{\partial^2 f^\lambda}{\partial x^\mu \partial x^\nu}
 +\bib{f^\lambda}{x^\rho}{\varGamma^\rho}_{\mu\nu}
 -\bib{f^\rho}{x^\mu}{\varGamma^\lambda}_{\rho\nu}
 -\bib{f^\rho}{x^\nu}{\varGamma^\lambda}_{\mu\rho}
 \right)
 \bib{}{{\varGamma^\lambda}_{\mu\nu}}, 
\end{align}
The energy-momentum current (\ref{palacurr}) is invariant 
with respect to the above gauge transformation.
Despite its non-trivial appearance, the energy-momentum current of
Palatini gravity eventually vanishes as mentioned at the beginning
of this section.

\section{Discussions}

We have constructed the theory of 
reparameterisation invariant
Lagrangian formulation in the setting of Kawaguchi geometry,
i.e. geometrisation of the variational principle for field theories,
and considered its application to several 
concrete models of field theory. In this formulation, we have shown that 
the conservation law of the energy-momentum currents appear as a part of the 
Euler-Lagrange equations. 
Mathematically, this result is due to the fact that 
the Kawaguchi manifold
is an extended configuration space 
including spacetime, 
and therefore the spacetime coordinates becomes cyclic for the 
field theory that usually does not have explicit dependency on spacetime coordinates.
Physically, the
Kawaguchi Lagrangian formulation tells us that the conservation law of 
energy-momentum currents
and the conventional equations of motions are on equal level. 
For example, instead of considering the Maxwell equations or Einstein's field equations, 
we can start the same discussion 
by choosing the conservation law.
One particular advantage
of this formulation is that we were able to propose a new way 
of understanding the 
energy-momentum conservation law of general relativity.
As in the case of other 
conventional
field theories, it is derived as a part of Euler-Lagrange equations,
as a result of existing cyclic coordinates. 
In the previous studies, energy-momentum currents of general relativity was 
defined as a pseudo-tensor~\cite{Bergmann-Thomson, Landau-Lifshitz, Dubois-Violette-Madore}, 
dependent only on the first order derivatives of $g^{\mu\nu}$, 
but in our result, they are derived as geometric quantities 
including second order derivatives of $g^{\mu\nu}$. 
In Einstein theory, the energy-momentum current of gravity becomes zero under on-shell condition: 
$\tilde{J}_\mu \stackrel{\sigma}{=} 0$.
However, it has a property of   on-shell gauge invariance,
${\cal L}_{\cal G} \tilde{J}_\mu \stackrel{\sigma}{=} 0$.
There exists various definitions and interpretations 
for energy-momentum current of gravity~\cite{Nester1999, Szabados}. 
While many of them do not satisfy gauge-invariance, 
our definition is gauge-invariant and reparameterisation invariant. 
Therefore,
it is a physical observable in real spacetime,
even if it vanishes. 

The Lagrangian formulation using Finsler/Kawaguchi manifold also simplified the Noether's theorem. 
The symmetry of the system is exactly the symmetry of the Lagrangian, which was the 
metric of Finsler/Kawaguchi manifold. 
And since our Finsler/Kawaguchi manifold encapsulate both spacetime and fields equivalently, 
no fibered structure was required and consequently the description of symmetries were simplified, namely, 
no distinction such as inner and exterior symmetries were introduced.
This is a notable calculational efficiency. 
For the future developments, we propose that Kawaguchi-Lagrangian formulation 
is a good candidate to construct models such as irreversible process and system which shows a hysteresis,
owing to the nature of Finsler and Kawaguchi geometry. 
Also, for Finsler-Lagrangian case, 
change of time coordinate generates a non-perturbative transformation such that
a harmonic oscillator turns into a free particle~\cite{OT2}.
Similarly, we expect that spacetime coordinate change in Kawaguchi-Lagrangian formulation
would offer us a non-perturbative transformation for field theory.

In the standard formulation, there are mainly two approaches to deal with the field theory; 
to consider the infinite dimensional configuration space and construct formal expressions, 
or to consider finite dimensional configuration space
with jet bundle structure. 
The first expression is simple but concrete problems are difficult to handle, 
and the second is applicable to concrete problems, but the structures and notations maybe 
sometimes difficult to handle for physicists. 
Our formulation is in a sense, a mixture of both, 
which has the simplicity of the former and the applicability of the latter. 
The actual calculations for concrete problems are accessible 
for most physicists as we have shown in the examples, 
and we hope this formulation could be helpful to 
understanding both past and future problems of physics.

\begin{acknowledgments}
We thank Lajos Tam\'{a}ssy, L\'{a}szl\'{o} Kozma, Masahiro Morikawa, Ken-ichi Nakao and Akio Sugamoto a for creative discussions. 
T. Ootsuka and E. Tanaka thank JSPS Institutional Program for Young Researcher Overseas Visits. 
E. Tanaka thanks SAIA grant and Yukawa Institute Computer Facility. 
M. Ishida acknowledges the grant-in-aid KAKENHI 25400272. 
This work was greatly inspired by late Yasutaka Suzuki.  
\end{acknowledgments}

\section*{Appendix A}

In the covariant Lagrangian formulation, we used frequently the sign $\stackrel{\sigma}{=}$,
which means the equality on the $4$-dimensional submanifold embedded in $M$.

With this symbol we mean,
\begin{eqnarray}
 A(x,dx) \stackrel{\sigma}{=} B(x,dx) \quad
 \Leftrightarrow \quad 
 \sigma^\ast A(x,dx)=\sigma^\ast B(x,dx),
\end{eqnarray}
where $A$ and $B$ are the functions of $x^\mu$ and 
$dx^{\mu_{i_1}\mu_{i_2}\cdots \mu_{i_k}} \, (1\leq k \leq N)$ and $\mathrm{dim} M=N$.

It is related to the ambiguity of the notations such as 
$dx^{\mu\nu\rho\sigma}$ and $dx^{\alpha\beta\gamma}\w d^2 x^{\mu\nu\rho\sigma}$.
The pull back of these quantities by parameterisation 
{$\sigma:=\sigma(s)$} is defined by,
\begin{align}
 &\sigma^\ast dx^{\mu\nu\rho\sigma}
 =\frac{\partial(x^\mu,x^\nu,x^\rho,x^\sigma)}{\partial (s^0,s^1,s^2,s^3)}
 ds^{0123}, \\
 &\sigma^\ast dx^{\alpha\beta\gamma}\w d^2 x^{\mu\nu\rho\sigma}
 =\frac{\partial \left(x^\alpha,x^\beta,x^\gamma, 
 \frac{\partial (x^\mu,x^\nu,x^\rho,x^\sigma)}{\partial (s^0,s^1,s^2,s^3)}\right)}
 {\partial (s^0,s^1,s^2,s^3)} \left(ds^{0123}\right)^2.
\end{align}
If we treat these variables always by its pull back as above, 
no ambiguity will enter in the formulae. 
However, we also used them as first and second order differential forms on $M$. 
For instance, Lie derivative ${\cal L}_X$ is defined by,
\begin{align}
 {\cal L}_X dx^{\mu\nu\rho\sigma}&=({\cal L}_Xdx^\mu)\w dx^{\nu\rho\sigma}
 -({\cal L}_X dx^\nu) \w dx^{\mu\rho\sigma}
 +({\cal L}_X dx^\rho) \w dx^{\mu\nu\sigma}
 -({\cal L}_X dx^\sigma) \w dx^{\mu\nu\rho} \nonumber \\
 &= dX^\mu \w dx^{\nu\rho\sigma}-dX^\nu \w dx^{\mu\rho\sigma}
 +dX^\rho \w dx^{\mu\nu\sigma}
 -dX^\sigma \w dx^{\mu\nu\rho}.
\end{align}
Namely, we considered $dx^{\mu\nu\rho\sigma}$ as a $4$-form on $M$, 
rather than the coordinate function on $\Lambda^4 TM$.
The meaning of the higher order differential form is not something new but notational. 
As we treat $dx^{\mu\nu\rho\sigma}$ as $4$-form (first order) on $M$, 
it acts on a $4$-vector field  
$v=\frac{1}{4!}v^{\alpha\beta\gamma\delta}\bib{}{x^\alpha}\w\bib{}{x^\beta}\w \bib{}{x^\gamma}\w \bib{}{x^\delta}$ over $M$,
which we define its action as, 
\begin{align}
 dx^{\mu\nu\rho\sigma}(v):=v^{\mu\nu\rho\sigma},
\end{align}
and we define the notation of the second order differential form 
$dx^{\alpha\beta\gamma}\w d^2 x^{\mu\nu\rho\sigma}$ 
by a recursive action of this first order form,  
\begin{align}
 dx^{\alpha\beta\gamma}\w d^2 x^{\mu\nu\rho\sigma}(v)
 &=\left\{
 dx^{\alpha\beta\gamma}\w d \left(dx^{\mu\nu\rho\sigma}(v) \right) 
 \right\}(v) \nonumber  \\
 &=dx^{\alpha\beta\gamma} \w dv^{\mu\nu\rho\sigma} (v)
 =v^{\alpha\beta\gamma\tau} \bib{v^{\mu\nu\rho\sigma}}{x^\tau}. 
\end{align}
Such operation allows us to simplify the calculation (such as taking the variation 
of the Kawaguchi metric) by using the standard computation technique of exterior and Lie derivative, 
without being aware of further details such as the background mathematical structures. 
While given the $4$-dimensional submanifold, a $4$-vector field could be defined 
as an oriented surface element on each points of $M$, 
the converse is not always true. 
This problem of the integrability of the vector field is the source of the ambiguity. 
Namely, the formula (such as $K$, ${\cal L}_X K$) expressed by variables on $M$, 
when pulled back to the $4$-dimensional integral submanifold, 
may give the same value for different expressions. 
For example, there are identities such as, 
\begin{align}
 \sigma^\ast dx^{\alpha\beta\gamma[\delta}dx^{\mu\nu\rho\sigma]}=0, \nonumber  \\
 \sigma^\ast dx^{\alpha\beta[\gamma}\w d^2 x^{\mu\nu\rho\sigma]}=0. \label{eq_dep}
\end{align}
Nevertheless, variational principle is given by the pull back equation,
${{\sigma}}^\ast \delta K=0$, which as we mentioned previously, 
does not include such ambiguity, and knowing that this pull back by ${\sigma}$ removes the ambiguity, 
we can safely use the symbol $\stackrel{\sigma}{=}$ to indicate 
the equivalence implied under the relation (\ref{eq_dep}).

\section*{Appendix B}

In this appendix, we will derive the first order energy momentum current (\ref{J_1st}) 
from the first order Kawaguchi metric of general relativity (\ref{1st-schwarz}),
\begin{align}
 K_{\rm 1st.}=
 -\frac{1}{4\kappa} \varepsilon_{\mu\nu\rho\sigma}\sqrt{-g}
 g^{\nu\xi}\tilde{\varGamma}^\mu{}_\lambda \w \tilde{\varGamma}^\lambda{}_\xi
 \w dx^{\rho\sigma}. \nonumber
\end{align}
The variation of $K_{\rm 1st.}$ with respect to $x^\mu$, 
$\delta_x K_{\rm 1st.}$, is calculated as follows:
\begin{align}
 \delta_x K_{\rm 1st.}
 =&
 -\frac{1}{4\kappa}
 \left\{
 \varepsilon_{\lambda\nu\rho\sigma}\sqrt{-g}g^{\nu\xi}
 \delta \left( {\tilde{\varGamma}{}^\lambda}_{\mu}\right)
 -\varepsilon_{\mu\nu\rho\sigma} \sqrt{-g} g^{\nu\lambda}
 \delta \left( {\tilde{\varGamma}{}^\xi}_{\lambda}\right)
 \right\} \w  dx^{\rho\sigma} \w {\tilde{\varGamma}{}^\mu}_{\xi} \nonumber
 \\
 &
  -\frac{1}{4\kappa}
 \varepsilon_{\mu\nu\rho\sigma}\sqrt{-g}g^{\nu\xi}
 {\tilde{\varGamma}{}^\mu}_{\lambda}\w  {\tilde{\varGamma}{}^\lambda}_{\xi}
 \w \delta \left(dx^{\rho\sigma}\right) \nonumber
 \\
  =&
 -\frac{1}{4\kappa} \varepsilon_{\mu\nu\rho\sigma} 
 d\left(\sqrt{-g}g^{\nu\xi} \right) \w \delta \left(dx^{\rho\sigma}\right)
 \w  {\tilde{\varGamma}{}^\mu}_{\xi} \nonumber
 \\
 &+\frac{1}{4\kappa} 
 \left(\varepsilon_{\lambda\nu\rho\sigma}\sqrt{-g}g^{\nu\xi}
  {\tilde{\varGamma}{}^\lambda}_{\mu}
 -\varepsilon_{\mu\nu\rho\sigma}\sqrt{-g}g^{\nu\lambda}
  {\tilde{\varGamma}{}^\xi}_{\lambda}
 \right) \w \delta \left(dx^{\rho\sigma}\right) \w  {\tilde{\varGamma}{}^\mu}_{\xi} \nonumber
 \\
 &
  -\frac{1}{4\kappa}
 \varepsilon_{\mu\nu\rho\sigma}\sqrt{-g}g^{\nu\xi}
 {\tilde{\varGamma}{}^\mu}_{\lambda}\w  {\tilde{\varGamma}{}^\lambda}_{\xi}
 \w \delta \left(dx^{\rho\sigma}\right) \nonumber
 \\
 =&
 d\left\{
  \frac{1}{2\kappa} \varepsilon_{\mu\nu\rho\sigma} 
 d\left(\sqrt{-g}g^{\nu\xi} \right) 
 \w  {\tilde{\varGamma}{}^\mu}_{\xi}\w dx^\rho \delta x^\sigma 
 -\frac{1}{2\kappa} 
  \varepsilon_{\mu\nu\rho\sigma}\sqrt{-g}g^{\nu\xi}
  {\tilde{\varGamma}{}^\mu}_{\lambda}
  \w {\tilde{\varGamma}{}^\lambda}_{\xi}
  \w dx^\rho \delta x^\sigma
 \right\} \nonumber
 \\
 &
 +d\left\{
  -\frac{1}{2\kappa} \varepsilon_{\mu\nu\rho\sigma} 
 d\left(\sqrt{-g}g^{\nu\xi} \right) 
 \w {\tilde{\varGamma}{}^\mu}_{\xi}\w dx^\rho
 +\frac{1}{2\kappa} 
  \varepsilon_{\mu\nu\rho\sigma}\sqrt{-g}g^{\nu\xi}
  {\tilde{\varGamma}{}^\mu}_{\lambda}
  \w {\tilde{\varGamma}{}^\lambda}_{\xi}
  \w dx^\rho
 \right\} \delta x^\sigma.
\end{align}
For the second equality, we used the variation of the relation (\ref{G-term}), {\em i.e.,
}
\begin{align}
\varepsilon_{\mu\nu\rho\sigma}d\left(\sqrt{-g}g^{\nu\xi}\right) \w 
  \delta \left(dx^{\rho\sigma}\right)
 \stackrel{\sigma}{=}&
 \left\{
 \varepsilon_{\lambda\nu\rho\sigma}\sqrt{-g}g^{\nu\xi}
  \delta \left(
  {\tilde{\varGamma}{}^\lambda}_\mu 
 \right)  
 -\varepsilon_{\mu\nu\rho\sigma} \sqrt{-g} g^{\nu\lambda}
 \delta \left(
 {\tilde{\varGamma}{}^\xi}_\lambda
 \right)
 \right\}
  \w dx^{\rho\sigma} \nonumber
 \\
 &
 + \left(
 \varepsilon_{\lambda\nu\rho\sigma}\sqrt{-g}g^{\nu\xi}
 {\tilde{\varGamma}{}^\lambda}_\mu 
 -\varepsilon_{\mu\nu\rho\sigma} \sqrt{-g} g^{\nu\lambda}
 {\tilde{\varGamma}{}^\xi}_\lambda 
 \right)
 \w \delta \left( dx^{\rho\sigma}\right).
\end{align}
Then we have the first order energy momentum current (\ref{J_1st})
\begin{align}
 \tilde{J}^{\rm 1st.}_\sigma=d\left(
 \frac{1}{2\kappa}\varepsilon_{\mu\nu\rho\sigma}\sqrt{-g}
 g^{\nu\xi}dx^{\rho}
 \right) \w \tilde{\varGamma}^\mu{}_\xi
 +
 \frac{1}{2\kappa}\varepsilon_{\mu\nu\rho\sigma}\sqrt{-g}g^{\nu\xi}
 \tilde{\varGamma}^\mu{}_\lambda \w \tilde{\varGamma}^\lambda{}_\xi \w dx^\rho. \nonumber
\end{align}

The consistency with the second order energy momentum current (\ref{e-current1}) 
can be also checked by,
\begin{align}
 \tilde{J}^G_\sigma&=\frac{1}{2\kappa}\varepsilon_{\mu\nu\rho\sigma}\sqrt{-g}g^{\nu\xi}
 {\tilde{R}{}^\mu}_\xi \w dx^\rho \nonumber
 \\
 =&
 d\left\{
 \frac{1}{2\kappa}\varepsilon_{\mu\nu\rho\sigma}\sqrt{-g}g^{\nu\xi}
 {\tilde{\varGamma}{}^\mu}_\xi \w dx^\rho
 \right\}
 -d \left(
 \frac{1}{2\kappa}\varepsilon_{\mu\nu\rho\sigma}\sqrt{-g}g^{\nu\xi}
 \right) \w {\tilde{\varGamma}{}^\mu}_\xi \w dx^\rho \nonumber
 \\
 &+\frac{1}{2\kappa}\varepsilon_{\mu\nu\rho\sigma}\sqrt{-g}g^{\nu\xi}
 {\tilde{\varGamma}{}^\mu}_\lambda \w {\tilde{\varGamma}{}^\lambda}_\xi \w dx^\rho \nonumber
 \\
 =&
 \tilde{J}^{\rm 1st.}_\sigma+ d\left\{
 \frac{1}{2\kappa}\varepsilon_{\mu\nu\rho\sigma}\sqrt{-g}g^{\nu\xi}
 {\tilde{\varGamma}{}^\mu}_\xi \w dx^\rho
 \right\}.
\end{align}


\end{document}